\documentclass[journal, twocolumn]{IEEEtran}
\IEEEoverridecommandlockouts

\usepackage[nocompress]{cite}
\usepackage{amssymb,amsfonts, amsthm}
\usepackage{graphicx}\graphicspath{{Figures/}}
\usepackage{epstopdf}
\usepackage{textcomp}
\usepackage{bbm}
\usepackage{xcolor}
\usepackage{algorithm}
\usepackage{algorithmicx}
\usepackage{algpseudocode}
\usepackage{verbatim}
\usepackage{subcaption}

\usepackage{algorithm}
\usepackage{multirow}
\usepackage{tabularx}
\usepackage{lipsum}
\usepackage{epsfig}
\usepackage{amsmath}
\usepackage{mathrsfs}
\usepackage{kotex}
\usepackage[font=small]{caption}
\usepackage{array}
\usepackage{hyperref}
\usepackage[table,dvipsnames]{xcolor}
\hypersetup{colorlinks,citecolor= red,filecolor= blue,linkcolor= blue,urlcolor=blue}
\usepackage{dblfloatfix}
\usepackage{siunitx}
\usepackage[normalem]{ulem}

\newcolumntype{L}[1]{>{\raggedright\let\newline\\\arraybackslash\hspace{0pt}}m{#1}}
\newcolumntype{C}[1]{>{\centering\let\newline\\\arraybackslash\hspace{0pt}}m{#1}}
\newcolumntype{R}[1]{>{\raggedleft\let\newline\\\arraybackslash\hspace{0pt}}m{#1}}
\def\multiset#1#2{\ensuremath{\left(\kern-.3em\left(\genfrac{}{}{0pt}{}{#1}{#2}\right)\kern-.3em\right)}}

\providecommand{\keywords}[1]{\textbf{\textit{Index Terms---}}\textbf{#1}}

\newcommand{\argmax}{\mathop{\mathrm{argmax}}}

\newtheorem{remark}{Remark}

\newtheorem{lemma}{Lemma}
\def\BibTeX{{\rm B\kern-.05em{\sc i\kern-.025em b}\kern-.08em
    T\kern-.1667em\lower.7ex\hbox{E}\kern-.125emX}}
\usepackage[printonlyused,withpage]{acronym}
\acrodef{RF}[RF]{radio-frequency}
\acrodef{mmWave}[mmWave]{millimeter wave}
\acrodef{THz}[THz]{terahertz}
\acrodef{UE}[UE]{user equipment}
\acrodef{BS}[BS]{base station}
\acrodef{LoS}[LoS]{line-of-sight}
\acrodef{NLoS}[NLoS]{non-line-of-sight}
\acrodef{SNR}[SNR]{signal-to-noise ratio}
\acrodef{SINR}[SINR]{signal-to-interference-plus-noise ratio}
\acrodef{MIMO}[MIMO]{multiple-input multiple-output}
\acrodef{MISO}[MISO]{multiple-input single-output}
\acrodef{SIMO}[SIMO]{single-input multiple-output}
\acrodef{OFDM}[OFDM]{orthogonal frequency division multiplexing}
\acrodef{UPA}[UPA]{uniform planar array}
\acrodef{AoA}[AoA]{azimuth angle of arrival}
\acrodef{ZoA}[ZoA]{zenith angle of arrival}
\acrodef{AoD}[AoD]{azimuth angle of departure}
\acrodef{ZoD}[ZoD]{zenith angle of departure}
\acrodef{CSI}[CSI]{channel state information}
\acrodef{LTE}[LTE]{Long-Term Evolution}
\acrodef{NR}[NR]{New Radio}
\acrodef{QoS}[QoS]{quality of service}
\acrodef{CE}[CE]{channel estimation}
\acrodef{ISAC}[ISAC]{integrated sensing and communications}
\acrodef{NMSE}[NMSE]{normalized mean square error}
\acrodef{RMSE}[RMSE]{root mean square error}
\acrodef{RB}[RB]{resource block}
\acrodef{LSFC}[LSFC]{large-scale fading coefficient}
\acrodef{SSFC}[SSFC]{small-scale fading coefficient}

\newacro{LMM-EMM}[LMM-EMM]{large multimodal model-based environment-aware mobility management}
\acrodef{UDN}[UDN]{ultra-dense network}
\acrodef{SBS}[SBS]{small base station}
\acrodef{MBS}[MBS]{macro base station}
\acrodef{RSRP}[RSRP]{reference signal received power}
\acrodef{S-SBS}[S-SBS]{serving SBS}
\acrodef{T-SBS}[T-SBS]{target SBS}

\acrodef{LLM}[LLM]{large language model}
\acrodef{LMM}[LMM]{large multimodal model}
\acrodef{LM}[LM]{language model}
\acrodef{AI}[AI]{artificial intelligence}
\acrodef{ML}[ML]{machine learning}
\acrodef{DL}[DL]{deep learning}
\acrodef{NLP}[NLP]{natural language processing}
\acrodef{DRL}[DRL]{deep reinforcement learning}
\acrodef{OD}[OD]{object detection}
\acrodef{DNN}[DNN]{deep neural network}
\acrodef{CNN}[CNN]{convolutional neural network}
\acrodef{LSTM}[LSTM]{long short-term memory}
\acrodef{KNN}[KNN]{K-nearest neighbor}
\acrodef{RNN}[RNN]{recurrent neural network}
\acrodef{LoRA}[LoRA]{low-rank adaptation}
\acrodef{CV}[CV]{computer vision}
\acrodef{GRU}[GRU]{gated recurrent unit}
\acrodef{SFT}[SFT]{supervised fine-tuning}
\acrodef{PEFT}[PEFT]{parameter-efficient fine-tuning}
\acrodef{NLL}[NLL]{negative log-likelihood}
\acrodef{DQN}[DQN]{deep Q-learning}

\acrodef{GCS}[GCS]{global coordinate system}
\acrodef{CCS}[CCS]{camera coordinate system}
\acrodef{CoMP}[CoMP]{coordinated multipoint}
\acrodef{BEV}[BEV]{bird's eye-view}
\acrodef{FoV}[FoV]{field of view}
\acrodef{DP}[DP]{dynamic programming}
\acrodef{CKM}[CKM]{channel knowledge map}
\acrodef{CCM}[CCM]{channel capacity map}
\acrodef{ECL}[ECL]{environment-aware channel capacity learning}
\acrodef{RRC}[RRC]{radio resource control}
\acrodef{AMF}[AMF]{access and mobility management function}
\acrodef{REM}[REM]{radio environment map}
\acrodef{GNSS}[GNSS]{global navigation satellite system}

\begin{document}
\title{Large Multimodal Model-Based \\ Environment-Aware Mobility Management}

\markboth{IEEE Transactions on Wireless Communications, Vol. XX, No. YY, ZZZZZZ 2026}{}

\author{
    Seokhyun~Jeong,~\IEEEmembership{Student Member,~IEEE},
    Sangmok~Shin,~\IEEEmembership{Student Member,~IEEE},
    Seungnyun~Kim,~\IEEEmembership{Member,~IEEE}, 
    Jiao~Wu,~\IEEEmembership{Member,~IEEE}, 
    and
    Byonghyo~Shim,~\IEEEmembership{Fellow,~IEEE}

    \thanks{
    Received 23 September, 2025; 
	revised 13 March, 2026 and 17 May, 2026;
    accepted 3 July, 2026.
    This work was supported in part by the National Research Foundation (NRF) of Korea under Grant RS-2022-NR070834 and
    2022M3C1A3099336, the Institute of Information \& Communications Technology Planning \& Evaluation(IITP)-ITRC(Information Technology Research Center) grant funded by the Korea government(MSIT)(IITP-2026-2021-0-02048) and the National Research Foundation, Singapore and Infocomm Media Development Authority under its Communications and Connectivity Bridging Funding Initiative.
    An earlier version of this paper was presented in part at the IEEE Global Communications Conference (GLOBECOM), Taipei, Taiwan, December 2025.
    (\textit{Corresponding author: Byonghyo~Shim.})
    }
    \thanks{
        Seokhyun~Jeong, Sangmok~Shin, and Byonghyo~Shim are with
		the Department of Electrical and Computer Engineering and the Institute of New Media and Communications,
		Seoul National University, 
		Seoul 08826
		Republic of Korea 
		(e-mail: {shjeong@islab.snu.ac.kr; smshin@islab.snu.ac.kr; bshim@snu.ac.kr}).
            
    	Seungnyun~Kim is with 
        the Information Systems Technology and Design (ISTD) pillar, Singapore University of Technology and Design, Singapore 487372
        (e-mail: {snkim94@mit.edu}).

        Jiao~Wu is with
        the Computer, Electrical and Mathematical Sciences and Engineering Division,
        King Abdullah University of Science and Technology, Thuwal 23955-6900 
        Saudi Arabia 
        (e-mail: jiao.wu@kaust.edu.sa).
	}	
}

\maketitle
\newcommand{\main}{2}
\newcommand{\blue}{black}
\newcommand{\bluetwo}{black}
\newcommand{\ad}[1]{\textcolor{\blue}{#1}}
\newcommand{\adtwo}[1]{\textcolor{\bluetwo}{#1}}
\newcommand{\graphsize}{1}

\begin{abstract}
Recently, \acp{LLM} have been successfully adopted in various fields, including wireless communications, robotics, and autonomous vehicles, owing to their outstanding adaptability and reasoning abilities.
Despite their huge potential, the application of \acp{LLM} for mobility management is relatively scarce since it requires not only analyzing wireless measurements but also predicting dynamic user trajectories and making real-time handover decisions across densely deployed \acp{SBS}.
In this paper, we propose an environment-aware mobility management scheme based on \acp{LMM}, which extend capabilities of \acp{LLM} to process multimodal sensing data.
By leveraging \acp{LMM}, the proposed scheme extracts contextual information on the surrounding environments from RGB-D images to capture \ac{UE} mobility patterns and identify signal reflections and blockages caused by static reflectors and dynamic obstacles.
Using the extracted environmental information, the proposed scheme learns the intrinsic mapping from \ac{UE} and \ac{SBS} positions to channel capacity, referred to as \ac{CCM}, from which future channel capacities along \ac{UE} trajectories are predicted. 
Based on the predicted channel capacities, we determine proactive handover decisions maximizing the cumulative channel capacities.
Simulation results demonstrate that the proposed scheme achieves substantial channel capacity improvements over conventional \ac{DL}-based approaches.
\end{abstract}


\acresetall	

\section{Introduction} \label{sec:I}
\Acp{LLM}, such as ChatGPT and Gemini, are attracting great attention these days for their powerful capabilities to solve unlimited complex tasks~\cite{brown2020language}.
Indeed, owing to the vast network size and extensive pretraining on diverse datasets, LLMs can solve a bewildering variety of problems with minimal examples (i.e., few-shot learning), or even without any example at all (i.e., zero-shot learning).
Benefited by advanced \ac{NLP} capabilities, \acp{LLM} can handle task instructions in the form of text prompts.
Recently, functionalities of \acp{LLM} have been extended to simultaneously process multimodal data, including images, audio, and video~\cite{liu2023visual}.
This extended version, called \ac{LMM}, can extract rich contextual information on the surrounding environment and thus can handle complex real-world tasks.
For this reason, \acp{LMM} can be used for a variety of applications, including smart factories, robotics, autonomous driving, and personal assistance~\cite{wei2022chain}.

\begin{figure}[t]
    \centering
    \ifthenelse{\equal{\main}{1}}
    {\newcommand{\mywidth}{0.7}}
    {\newcommand{\mywidth}{0.9}}
    \includegraphics[width=\mywidth\linewidth]{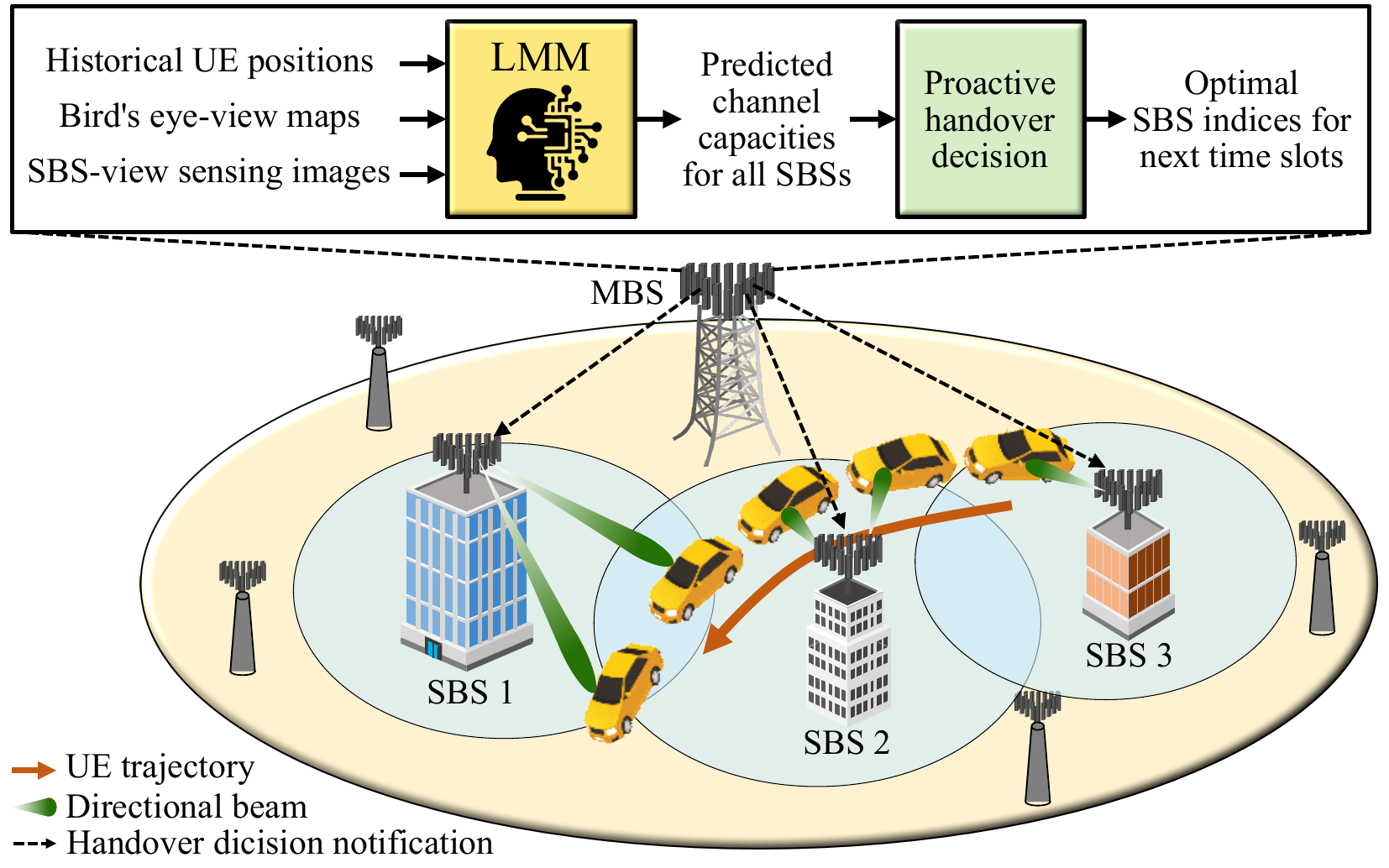}
    \caption{Visualization of mobility management in UDN systems.}
    \label{fig:1}
    \vspace{-1.2em}
\end{figure}

While there are recent research efforts integrating \ac{LMM} into wireless systems, papers addressing \ac{LMM}-based mobility management are relatively scarce~\cite{maatouk2024large, yang2026large}.
This is because mobility management involves far more diversified functionalities than those required by resource allocation, such as predicting the dynamic movements of \acp{UE} and solving intricate cell association problems.
In fact, the primary goal of mobility management is to associate each \ac{UE} with an appropriate \ac{BS} to ensure a reliable connection (i.e., cell association)~\cite{ahn2024sensing}.
In particular, in \acp{UDN}, wireless channels between a \ac{UE} and a \ac{SBS} fluctuate rapidly as the \ac{UE} moves through densely deployed \acp{SBS} (e.g., pico and femto cells) and thus a delay in the cell association process will affect the reliable connection considerably (see Fig.~\ref{fig:1})~\cite{moon2023energy, kim2023efficient, kamel2016ultra}.

\ad{To guarantee seamless connectivity in \acp{UDN}, a handover technique transferring the ongoing cell association to a nearby \ac{SBS} is needed.
In 4G \ac{LTE} and 5G \ac{NR}, handovers are triggered based on \ac{UE}-reported \ac{RSRP} measurements when the \ac{T-SBS} provides a higher \ac{RSRP} than the \ac{S-SBS} over four consecutive reports~\cite{3gpp38331}.
While this approach has been used for many years, it is not quite effective for 5G and upcoming 6G networks, since it induces a considerable handover delay~\cite{ahn2024sensing}\footnote{\ad{Note that the \ac{S-SBS} can initiate the handover only after receiving a sequence of \ac{RSRP} reports from the \ac{UE}.
In 5G \ac{NR}, the time interval between adjacent \ac{RSRP} reports is $40$-$120$\,ms, and thus the total handover latency can reach up to 360\,ms.
Clearly, this time far exceeds the channel coherence time of 5G \ac{NR}. For example, in 5G \ac{NR} with carrier frequency of $f_{\mathrm{c}}=3.5\,$GHz and \ac{UE} speed of $v=10\,$m/s, the large-scale fading channel coherence time is $40\sqrt{\frac{9}{16\pi}}\frac{c}{v f_{\mathrm{c}}} \approx 145.1$\,ms~\cite{clarke1968statistical}.}}.}
\ad{To mitigate this issue, several approaches for reducing the handover delay by proactively performing handover have been suggested.
In~\cite{yan2019machine, shah2022multi, park2024mobility}, proactive mobility management techniques that utilize \ac{ML} and \ac{DL} models have been studied.
In~\cite{jiao2021enabling,huang2022self, sun2025proactive}, \ac{DRL}-based mobility management techniques that use policy optimization for the handover control have been proposed.
The potential weakness of these techniques is that they rely exclusively on wireless measurements (e.g., \ac{RSRP}), which hinders rapid adaptation to abrupt channel variations caused by the dynamic environment (e.g., human/vehicle movements and sudden environmental changes)~\cite{kim2024role}.
Recently, \ac{ISAC}-based handover techniques that leverage the sensing data (e.g., images, radar signals, and LiDAR point clouds) to predict potential link blockages have gained attention~\cite{charan2021vision, demirhan2022radar, liu2023vision,chaccour2024joint,feng2025networked}.}
While the \ac{ISAC}-based approaches can incorporate environmental features to some extent, these techniques focus primarily on detecting \ac{LoS} path and overlook the impact of multipath propagation (e.g., reflection and scattering), limiting their effectiveness in rich scattering environments~\cite{shen2016non}.

An aim of this paper is to put forth an environment-aware mobility management framework empowered by the \ac{LMM} technology.
The proposed framework, referred to as \acfi{LMM-EMM}, preemptively associates the \ac{UE} with a proper \ac{SBS} by leveraging environmental context and \ac{UE} mobility patterns extracted from multimodal data (e.g., RGB-D images and wireless measurements).
Key ingredient in achieving this mission is \ac{LMM}, a generative \ac{AI} model specialized in extracting correlated features across multiple modalities and performing high-level reasoning.
In the proposed framework, \ac{LMM} learns the \ac{UE} movement pattern and the reflection geometry between the \ac{UE} and \acp{SBS} by analyzing the correlations between the physical environments (e.g., buildings, road structures, and dynamic obstacles) obtained from \ac{BEV} maps, \ac{SBS}-view sensing images, and also the wireless measurements.
These features are then exploited to construct a fundamental mapping between the \ac{UE} position and the channel capacity, henceforth dubbed as \acfi{CCM}.
Using \ac{CCM} along with the predicted \ac{UE} trajectory, the channel capacity can be estimated without requiring any real-time measurements, thereby facilitating fast and accurate environment-aware handovers.
The main contributions of this paper are summarized as follows:
\begin{itemize}
    \item We propose an \ac{LMM}-based environment-aware mobility management framework that ensures fast and reliable connectivity.
    Unlike conventional schemes that rely primarily on wireless signal measurements, LMM-EMM exploits the multimodal reasoning capability of \ac{LMM} to incorporate a richer contextual understanding of the surrounding environment, thereby supporting more predictive and robust mobility decisions.
    \item We develop an environment-aware channel capacity estimation technique that captures the reflection geometry of the propagation channel.
    To this end, we construct the \ac{CCM}, an end-to-end mapping from the \ac{UE} position to channel capacity.
    A main step of the \ac{CCM} construction is analyzing the environmental context and inferring the underlying propagation characteristics, and we use \ac{LMM} for this purpose.
    By leveraging \ac{CCM}, we can accurately predict the future channel capacity for each \ac{SBS} and make proactive handover decisions that reduce latency and enhance link reliability.
    \item From extensive simulations on realistic \ac{UDN} environments, we show that the proposed \ac{LMM-EMM} achieves a significant capacity gain over the conventional mobility management techniques.
    Specifically, \ac{LMM-EMM} achieves about 45\% improvement in channel capacity compared to the 5G NR mobility management scheme.
    Even when compared with the \ac{LSTM}-based and \ac{DRL}-based approaches, \ac{LMM-EMM} yields 21\% and 15\% capacity gains, respectively.
\end{itemize}

The rest of this paper is organized as follows:
In Section \ref{sec:II}, we explain the \ac{UDN} system model and briefly introduce conventional mobility management techniques. 
In Section \ref{sec:III}, we present the overall architecture of \ac{LMM-EMM}.
In Section \ref{sec:IV}, we discuss practical issues.
In Section \ref{sec:V}, we demonstrate the numerical results and conclude the paper in Section \ref{sec:VI}.

\textit{Notations}: 
Bold upper and lower case symbols denote matrices and vectors, respectively.
Sets are denoted by calligraphic font, for example, $\mathcal{X}$.
Superscript $(\cdot)^{\mathrm{T}}$ denotes the transpose.
$x^{(t)}$ and $x^{(t_1:t_2)}$ denote the value of $x$ at time slot $
t$ and $\{x^{(t_1)}, x^{(t_1+1)}, \cdots, x^{(t_2)}\}$ where $t_1 < t_2$, respectively.
$\mathbf{X}_1\otimes \mathbf{X}_2$ denotes the Kronecker product of matrices $\mathbf{X}_1$ and $\mathbf{X}_2$.
$||\mathbf{x}||_2$ and $[\mathbf{x}]_n$ denote the Euclidean norm and the $n$th element of a vector $\mathbf{x}$, respectively.
$\mathbb{E}[\cdot]$ denotes an expectation of a random variable.
$\vert \mathcal{X} \vert$ denotes the number of elements in a set $\mathcal{X}$.
$\bigcup_i \mathcal{X}_i$ denotes the union of the sets $\mathcal{X}_i$ over all $i$.

\section{Ultra-dense Network System}\label{sec:II}
In this section, we provide a brief overview of \ac{UDN} systems and discuss conventional mobility management techniques.

\subsection{Downlink UDN System Model}\label{sec:II.A}
We consider a \ac{mmWave} \ac{MISO} \ac{UDN} system consisting of $M$ \acp{SBS} equipped with $N=N_{x} \times N_{y}$ \ac{UPA} antennas and a \ac{UE} equipped with a single antenna.
The set of \ac{SBS} indices is denoted as $\mathcal{M}=\{1,2,\dotsc,M\}$.
The \acp{SBS} are connected to the \ac{MBS} through backhaul links to share transmit data and control signals.
We consider a \ac{GCS} where the position vectors of the $m$th \ac{SBS} and the \ac{UE} at the $t$th time slot are $\mathbf{p}_{\mathrm{sbs},m}^{(t)} = [x_{\mathrm{sbs},m} \,\, y_{\mathrm{sbs},m} \,\, z_{\mathrm{sbs},m}]^{\mathrm{T}}$ and $\mathbf{p}_{\mathrm{ue}}^{(t)} = \big[x_{\mathrm{ue}}^{(t)} \,\, y_{\mathrm{ue}}^{(t)} \,\,z_{\mathrm{ue}}^{(t)}\big]^{\mathrm{T}}$, respectively.
We also consider an \ac{OFDM} system with $S$ subcarriers, a carrier frequency of $f_{\mathrm{c}}$, and a system bandwidth of $B$.
The frequency of the $s$th subcarrier is $f_s = f_{\mathrm{c}} + (s-\frac{S}{2})\frac{B}{S}$ for $s = 1,\cdots, S$.
Assuming the equal power allocation among the data streams, the \ac{MISO} channel capacity $R^{(t)}(m)$ between the $m$th \ac{SBS} and the \ac{UE} at the time slot $t$ is
\vspace{-0.2em}
\begin{IEEEeqnarray}{rCl} \label{capacity}
    R^{(t)}(m) &=& B\sum_{s=1}^{S}\log_{2}\bigg(1+\frac{P_{\mathrm{t}}}{N\sigma_{\mathrm{n}}^{2}}\big\Vert\mathbf{h}_{m}^{(t)}[s]\big\Vert_{2}^{2}\bigg)
    \vspace{-0.2em}
\end{IEEEeqnarray}
\noindent where $\mathbf{h}_{m}^{(t)}[s]\in\mathbb{C}^{N}$ is the downlink channel vector from the $m$th \ac{SBS} to the \ac{UE} at the $s$th subcarrier and the $t$th time slot, $P_{\mathrm{t}}$ is the \ac{SBS} transmission power, and $\sigma_{\mathrm{n}}^{2}$ is the noise power~\cite{molisch2002capacity}.
Since $R^{(t)}(m)$ depends on $\{\mathbf{h}_{m}^{(t)}[s]\}_{s=1}^{S}$, it varies dynamically as the \ac{UE} moves~\cite{tang2024utilizing}. 
Thus, if the link quality of \ac{S-SBS} is lower than the predefined threshold due to the movement of the \ac{UE}, the \ac{UE} must search for a \ac{T-SBS} and initiate a handover process.

In our work, we adopt a block-fading geometric multipath channel model where the channel remains constant within a time slot with duration $\tau_{s}$.
Each time slot is divided into the \ac{CE} period $\tau_{\mathrm{ce}}$, during which \acp{RB} are dedicated to transmitting reference signals (e.g., CSI-RS), and the data transmission period $\tau_{\mathrm{dt}}$.
Under this channel model, the downlink channel vector $\mathbf{h}_{m}^{(t)}[s]$ is expressed as the sum of an \ac{LoS} path (denoted by $l=0$) and $L-1$ \ac{NLoS} paths:
\vspace{-0.2em}
\begin{IEEEeqnarray}{rCl} \label{channel}
    \mathbf{h}_{m}^{(t)}[s] &=& 
    \sqrt{\beta_{m,0}^{(t)}} 
    \alpha_{m,0}^{(t)} 
    e^{-j2\pi f_s \tau_{m,0}^{(t)}} 
    \mathbf{a}(\theta_{m,0}^{(t)}, \phi_{m,0}^{(t)})
    \ifthenelse{\equal{\main}{2}}
    {\notag \\ &&}{}
    + \sum_{l=1}^{L-1} 
    \sqrt{\beta_{m,l}^{(t)}} 
    \alpha_{m,l}^{(t)} 
    e^{-j2\pi f_s \tau_{m,l}^{(t)}} 
    \mathbf{a}(\theta_{m,l}^{(t)}, \phi_{m,l}^{(t)})
    \vspace{-0.2em}
\end{IEEEeqnarray}
where $\beta_{m,l}^{(t)}$ is the \ac{LSFC}, $\alpha_{m,l}^{(t)} \sim \mathcal{CN}(0, 1)$ is the \ac{SSFC}, $\tau_{m,l}^{(t)}$ is the time delay, and $\theta_{m,l}^{(t)} \in [0, 2\pi)$ and $\phi_{m,l}^{(t)} \in [0, \pi]$ are the \ac{AoD} and \ac{ZoD} of the $l$th path between the $m$th \ac{SBS} and the \ac{UE}, respectively.  
Also, $\mathbf{a}\big(\theta_{m,l}^{(t)}, \phi_{m,l}^{(t)}\big) \in \mathbb{C}^{N}$ is the \ac{SBS} array steering vector given by
\begin{IEEEeqnarray}{rCl}
    \mathbf{a}\big(\theta_{m,l}^{(t)}, \phi_{m,l}^{(t)}\big) &= 
    & \Big[1\, \cdots\,
    e^{j\frac{2\pi d}{\lambda} (N_{x} -1)\cos\theta_{m,l}^{(t)} \sin\phi_{m,l}^{(t)}}\Big] 
    \ifthenelse{\equal{\main}{2}}
    {\nonumber \\ && \,\,}{}
    \otimes
    \Big[1\, \cdots\,
    e^{j\frac{2\pi d}{\lambda} (N_{y} -1)\sin\theta_{m,l}^{(t)} \sin\phi_{m,l}^{(t)}}\Big]
    \IEEEeqnarraynumspace
\end{IEEEeqnarray}
where $d$ and $\lambda$ are the antenna spacing and signal wavelength, respectively.

\ad{
Specifically, $\beta_{m,l}[s]$ is expressed as
\begin{IEEEeqnarray}{rcl}
    \beta_{m,l}[s]&=&\frac{\Gamma_{m,l}}{4\pi f_{s} \tau_{m,l}} e^{-\Big(\frac{8 \pi^2 \sigma_{m,l}^2 f_{s}^2 \cos^2(\psi_{m,l})}{c^2}+\frac{c \tau_{m,l} k_{\mathrm{abs}}(f_s) }{2}\Big) }\!\IEEEeqnarraynumspace  
    \label{pathgain}
\end{IEEEeqnarray}
where $k_{\mathrm{abs}}(f_s)$, $c$, $\Gamma_{m,l}$, $\sigma_{m,l}$, and $\psi_{m,l}$ are the molecular absorption coefficient, the speed of light, the Fresnel coefficient, the roughness coefficient, and the angle of incidence, respectively~\cite{solomitckii2016characterizing}.
These parameters are determined by the \ac{UE} and \ac{SBS} positions $\mathbf{p}_{\mathrm{ue}}$ and $\mathbf{p}_{\mathrm{sbs},m}$, as well as the reflection surface $\mathcal{E}_{m,l}=\{\mathbf{x}\mid \mathbf{n}_{m,l}^{\mathrm{T}} \mathbf{x}=b_{m,l}\}$ ($\|\mathbf{n}_{m,l}\|_2=1$)~\cite{kim2025large}:
\begin{IEEEeqnarray}{rCl}
    \tau_{m,l} 
    &=& 
    \frac{1}{c}
    \| \mathbf{d}_m
    - 2(\gamma_{m,l} - b_{m,l}) \mathbf{n}_{m,l}
    \|_2
    \IEEEeqnarraynumspace \label{tau}\\
    \psi_{m,l}
    &=&
    \mathrm{arccos}
    \bigg(
    \frac{
    \vert \gamma_{m,l} 
    + \delta_{m,l} 
    \vert
    }{
        \sqrt{ 
        \| \mathbf{d}_m \|_2^2
        + 4\gamma_{m,l}\delta_{m,l}
        }
    }
    \bigg)
    \IEEEeqnarraynumspace \label{psi}\\
    \Gamma_{m,l} 
    &=& \frac{\cos \psi_{m,l} - \sqrt{\epsilon - \sin^2 \psi_{m,l}}}
    {\cos \psi_{m,l} - \sqrt{\epsilon + \sin^2 \psi_{m,l}}}\vspace{-0.2em} \label{Gamma}
\end{IEEEeqnarray}
where $\mathbf{d}_m = \mathbf{p}_{\mathrm{ue}} - \mathbf{p}_{\mathrm{sbs},m} $, $\gamma_{m,l} = \mathbf{n}_{m,l}^{\mathrm{T}}\mathbf{p}_{\mathrm{ue}}-b_{m,l}$, $\delta_{m,l}=\mathbf{n}_{m,l}^{\mathrm{T}}\mathbf{p}_{\mathrm{sbs},m}-b_{m,l}$, and $\epsilon$ is the dielectric permittivity of the reflector.}

By concatenating $\mathbf{h}_{m}^{(t)}[s]$ for all subcarriers, we obtain the frequency domain channel matrix $\mathbf{H}_{m}^{(t)}\in \mathbb{C}^{N \times S}$:
\begin{IEEEeqnarray}{rCl}
    \mathbf{H}_{m}^{(t)} &=& \big[\mathbf{h}_{m}^{(t)}[1]\, \mathbf{h}_{m}^{(t)}[2]\, \cdots \,\mathbf{h}_{m}^{(t)}[S]\big] \IEEEeqnarraynumspace \\
    &=& \sum_{l=0}^{L-1} \sqrt{\beta_{m,l}^{(t)}} \alpha_{m,l}^{(t)}\mathbf{a}\big(\theta_{m,l}^{(t)}, \phi_{m,l}^{(t)}\big) \mathbf{b}^{\mathrm{T}}\big(\tau_{m,l}^{(t)}\big)
    \IEEEeqnarraynumspace
\end{IEEEeqnarray}
where $\mathbf{b}(\tau_{m,l}^{(t)})\in\mathbb{C}^{S}$ is the phase shift vector of the \ac{OFDM} subcarriers given by
\begin{IEEEeqnarray}{rCl}
\mathbf{b}(\tau_{m,l}^{(t)}) = \Big[e^{-j 2\pi f_1 \tau_{m,l}^{(t)}} \  e^{-j 2\pi f_2 \tau_{m,l}^{(t)}} \  \cdots \  e^{-j 2\pi f_{S} \tau_{m,l}^{(t)}}\Big]^{\mathrm{T}}. \IEEEeqnarraynumspace
\end{IEEEeqnarray}
For brevity, we define the collection of the geometric channel parameters $\mathcal{G}_{m,l}^{(t)}$ of the $l$th path between the $m$th \ac{SBS} and \ac{UE} at the $t$th time slot as
\begin{equation}
    \mathcal{G}_{m,l}^{(t)} = \big\{\beta_{m,l}^{(t)}, \alpha_{m,l}^{(t)}, \tau_{m,l}^{(t)}, \vartheta_{m,l}^{(t)}, \varphi_{m,l}^{(t)}, \theta_{m,l}^{(t)}, \phi_{m,l}^{(t)}\big\}.
\end{equation}
Then, we can express $\mathbf{H}_{m}^{(t)}$ as a function of $\{\mathcal{G}_{m,l}^{(t)}\}_{l=1}^{L}$ as
\begin{IEEEeqnarray}{c} \label{channel_parameter}
    \mathbf{H}_{m}^{(t)} = \mathbf{P}\big(\mathcal{G}_{m,0}^{(t)}\big) + \sum_{l=1}^{L-1}\mathbf{P}\big(\mathcal{G}_{m,l}^{(t)}\big)
\end{IEEEeqnarray}
where $\mathbf{P}(\mathcal{G}_{m,l}^{(t)}) \in \mathbb{C}^{N\times S}$ denotes the $l$th multipath component of $\mathbf{H}_{m}^{(t)}$ corresponding to $\mathcal{G}_{m,l}^{(t)}$:
\begin{equation}
    \mathbf{P}(\mathcal{G}_{m,l}^{(t)}) 
    = \sqrt{\beta_{m,l}^{(t)}} 
    \alpha_{m,l}^{(t)}
    \mathbf{a}\big(\theta_{m,l}^{(t)}, \phi_{m,l}^{(t)}\big) \mathbf{b}^{\mathrm{T}}\big(\tau_{m,l}^{(t)}\big).
\end{equation}
Note that geometric channel parameters vary continuously when reflectors remain unchanged.
However, changes in reflectors can cause abrupt transitions, which lead to discontinuities in channel capacity~\cite{ju2024transformer}.

\begin{figure}[t]
    \centering
    \ifthenelse{\equal{\main}{1}}
    {\newcommand{\mywidth}{0.7}}
    {\newcommand{\mywidth}{0.95}}
    \includegraphics[width=\mywidth\linewidth]{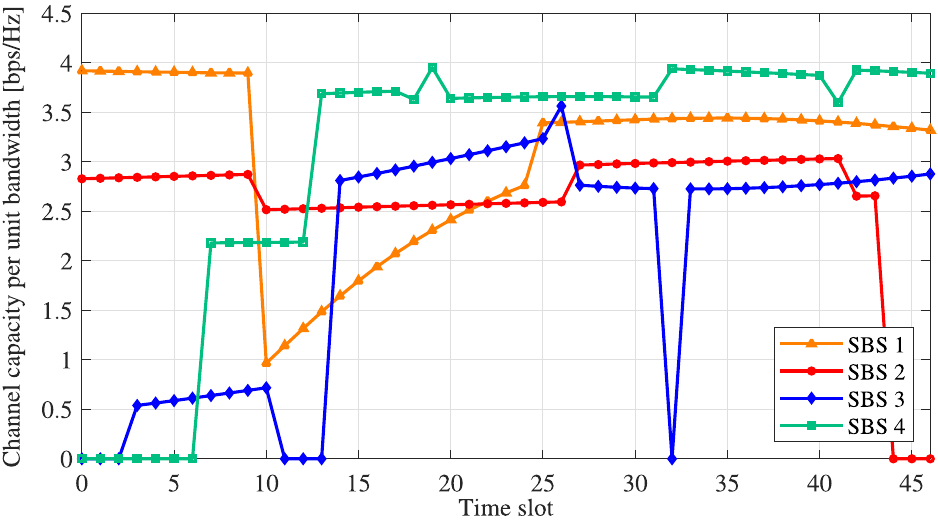}
    \caption{Piecewise continuity of channel capacity with \ac{UE} movement.}
    \label{fig:2}
    \vspace{-1em}
\end{figure}

\begin{remark} \label{remark1}
    As shown in Fig.~\ref{fig:2}, the channel capacity $R^{(t)}(m)$ for each \ac{SBS} is piecewise continuous over time.
    The discontinuity arises from the discontinuity of reflection points on different reflectors.
\end{remark}

{\color{\blue}
\subsection{Channel Capacity Map}
In this subsection, we introduce the concept of the \ac{CCM}, which characterizes the achievable channel capacity as a function of the \ac{UE} position, the \ac{SBS} position, and environmental factors (e.g., reflectors).
For notational simplicity, the time index $t$ is omitted without loss of generality.
\begin{lemma} \label{lemma:1}
    When the number of transmit antennas $N$ is sufficiently large, the channel capacity ${R}_{\mathrm{ideal}}(m)$ can be expressed as a function of \acp{LSFC} $\{\beta_{m,l}[s]\}_{l=0}^{L-1}$. That is,
    \begin{equation} \label{eqlemma:1}
         R_{\mathrm{ideal}}(m) = B\sum_{s=1}^{S}
         \log_{2} \bigg( 1+\frac{P_{\mathrm{t}}}
         {\sigma_{\mathrm{n}}^{2}} \bigg(\beta_{m,0}[s] +
         \sum_{l=1}^{L-1}\beta_{m,l}[s] \bigg)
         \bigg).
    \end{equation}
    \begin{proof}
        See Appendix~\hyperref[app:A]{A}.
    \end{proof}
\end{lemma}
\noindent In practice, moving obstacles (e.g., vehicles and pedestrians) can intermittently block the \ac{LoS} link between \ac{UE} and \ac{SBS}.
Such dynamic blockages significantly impact the achievable channel capacity due to the pronounced power disparity between the \ac{LoS} and \ac{NLoS} components~\cite{solomitckii2016characterizing}. 
\begin{remark}\label{remark2}
    The achievable channel capacity is a function of static channel capacity and \ac{LoS} indicator as
    \begin{equation}
        {R}(m)
        = c_{\mathrm{los}}(m)
        {R}_{\mathrm{ideal}}(m) + (1-c_{\mathrm{los}}(m))
        {R}_{\mathrm{nlos}}(m)
        \label{dynamic_capacity}
    \end{equation}
    where $R_{\mathrm{nlos}}(m)$ is the capacity of \ac{NLoS} channel given by
    \begin{equation}
         R_{\mathrm{nlos}}(m) = B\sum_{s=1}^{S}
         \log_{2} \bigg( 1+\frac{P_{\mathrm{t}}}
         {\sigma_{\mathrm{n}}^{2}}
         \sum_{l=1}^{L-1}\beta_{m,l}[s]
         \bigg). \label{nlos}
    \end{equation}
    Also, $c_{\mathrm{los}}(m)$ is the \ac{LoS} indicator defined as
    \begin{equation} \label{clos}
        c_{\mathrm{los}}(m)=
        \begin{cases}
            1 & \text{\!\!if the \ac{LoS} for \ac{SBS} $m$ exists}\\
            0 & \text{\!\!otherwise}.
        \end{cases}
    \end{equation}
\end{remark}

From Equation \eqref{eqlemma:1} and \eqref{nlos}, one can see that $R_{\mathrm{ideal}}(m)$ and $R_{\mathrm{nlos}}(m)$ are functions of \acp{LSFC} $\{\beta_{m,l}[s]\}_{l=0}^{L-1}$.
Moreover, as shown in Equation \eqref{pathgain}-\eqref{Gamma}, the \ac{LSFC} $\beta_{m,l}[s]$ is determined by the \ac{UE} position $\mathbf{p}_{\mathrm{ue}}$, \ac{SBS} position $\mathbf{p}_{\mathrm{sbs},m}$, and the reflection surface $\mathcal{E}_{m,l}=\{\mathbf{x}\mid \mathbf{n}_{m,l}^{\mathrm{T}} \mathbf{x}=b_{m,l}\}$.
Therefore, $R_{\mathrm{ideal}}(m)$ and $R_{\mathrm{nlos}}(m)$ can be directly obtained from $\mathbf{p}_{\mathrm{ue}}$, $\mathbf{p}_{\mathrm{sbs},m}$, and $\mathcal{E}=\cup_{m,l} \,\mathcal{E}_{m,l}$.

\begin{remark}\label{remark3}
    The static channel capacities ${R}_{\mathrm{ideal}}(m)$ and $ {R}_{\mathrm{nlos}}(m)$ are fully characterized by three components: the \ac{UE} position $\mathbf{p}_{\mathrm{ue}}$, the \ac{SBS} position $\mathbf{p}_{\mathrm{sbs},m}$, and the reflector information $\mathcal{E}$, through \ac{CCM} $f_{\mathrm{ccm}}$:
    \begin{IEEEeqnarray}{rCl}
        \big({R}_{\mathrm{ideal}}(m), {R}_{\mathrm{nlos}}(m)\big)&=& f_{\mathrm{ccm}}(\mathbf{p}_{\mathrm{ue}}, \mathbf{p}_{\mathrm{sbs},m}, 
        \mathcal{E}). \IEEEeqnarraynumspace \label{composition}
    \end{IEEEeqnarray}
\end{remark}

\noindent Remarks~\ref{remark2} and \ref{remark3} show that predicting $R(m)$ requires the \ac{UE} position $\mathbf{p}_{\mathrm{ue}}$, the \ac{SBS} positions $\{\mathbf{p}_{\mathrm{sbs},m} \mid m \in \mathcal{M} \}$, the reflector information $\mathcal{E}$, and the \ac{LoS} indicators $\{c_{\mathrm{los}}(m) \mid m \in \mathcal{M} \}$.
}

\begin{figure}[t]
    \centering
    \ifthenelse{\equal{\main}{1}}
    {\newcommand{\mywidth}{0.6}}
    {\newcommand{\mywidth}{0.9}}
    \includegraphics[width=\mywidth\linewidth]{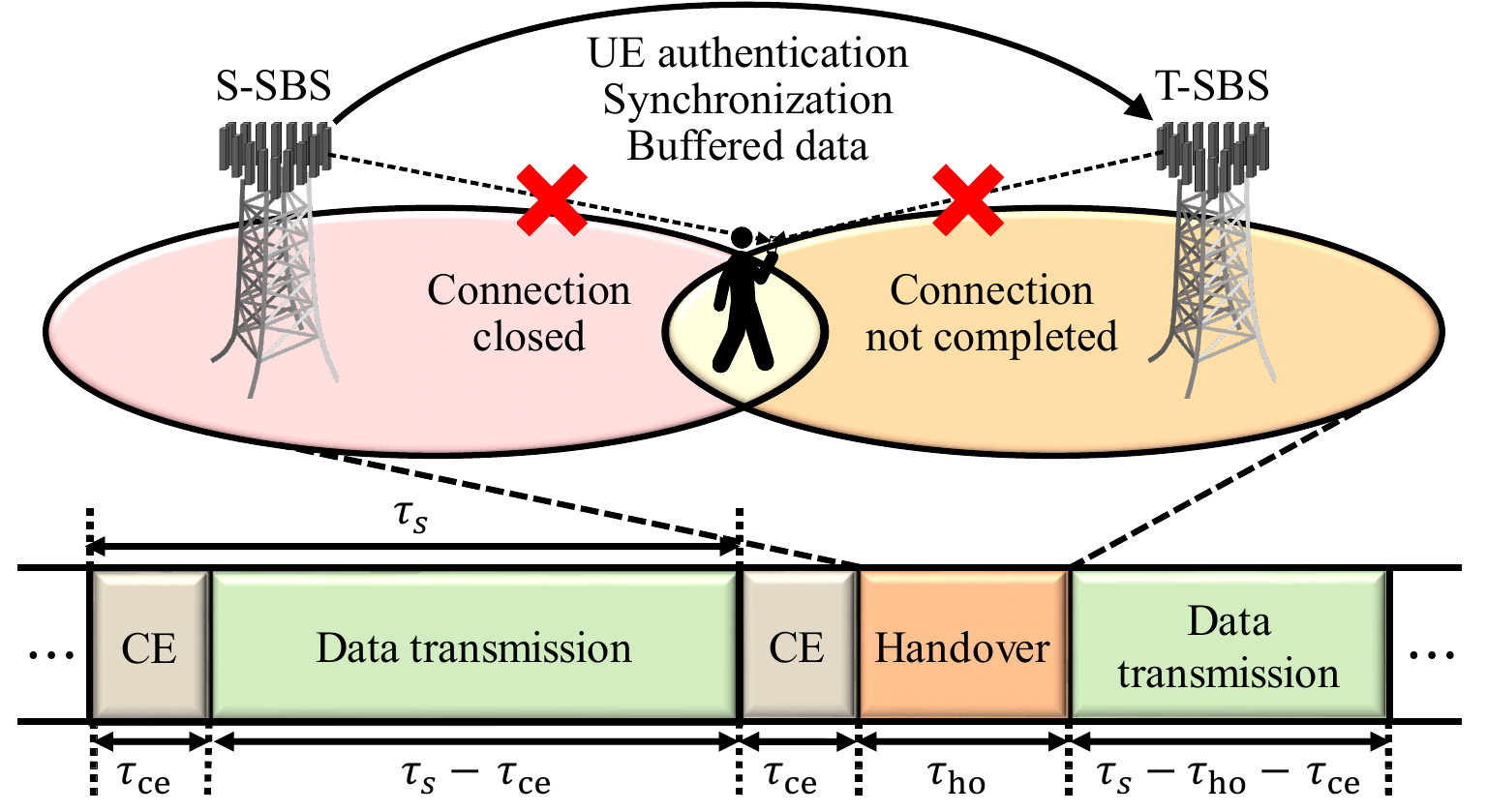}
    \caption{Illustration of service interruption during the handover.}
    \label{fig:3}
    \vspace{-1.2em}
\end{figure}

\subsection{Handover Process}\label{sec:II.B}
The handover process in 5G \ac{NR} consists of three main phases: preparation, execution, and completion~\cite{3gpp36423}. 
The preparation phase begins when the \ac{RSRP} of \ac{T-SBS} exceeds that of \ac{S-SBS} by a configured offset (i.e., event A3~\cite{3gpp36423}).
At this point, the \ac{UE} initiates the handover process by sending the \ac{RRC} measurement report to \ac{S-SBS}.
If \ac{S-SBS} continues to receive these reports throughout the designated period (i.e., time-to-trigger), it sends the handover request to \ac{T-SBS}, followed by a handover acknowledgment from \ac{T-SBS} to \ac{S-SBS}.
Next, in the execution phase, \ac{S-SBS} transfers the \ac{UE} identification and security information (e.g., encryption algorithms) to \ac{T-SBS}, along with any downlink data that has not yet been delivered to the \ac{UE}.
Then, the \ac{UE} establishes a new connection to \ac{T-SBS} via the random access.
Lastly, in the completion phase, the core network finalizes the handover process by redirecting the \ac{UE} data path to \ac{T-SBS}.

\begin{figure*}[t]
    \centering
    \includegraphics[width=1\linewidth]{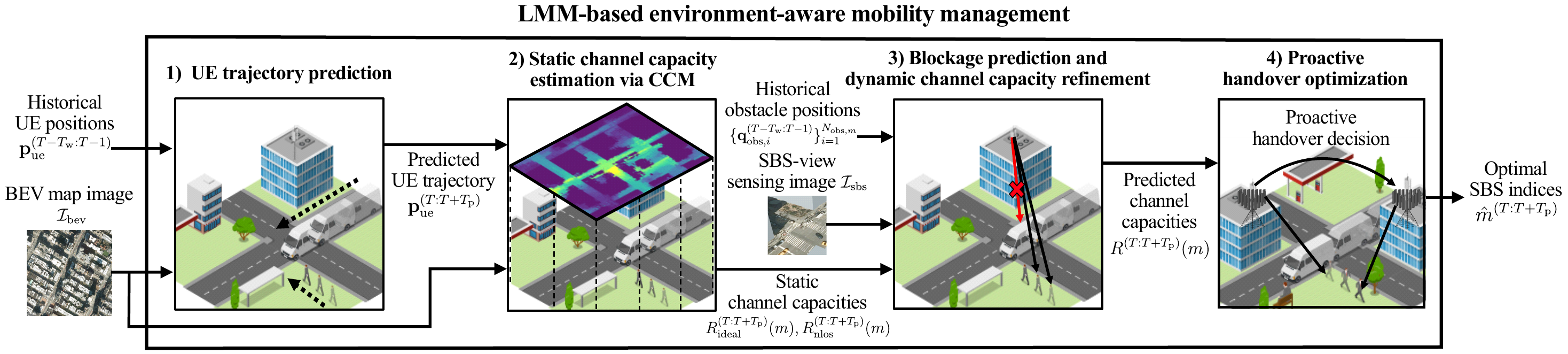}
    \caption{Overall procedure of the proposed \ac{LMM-EMM}.}
    \label{fig:4}
    \vspace{-1.5em}
\end{figure*}

Note that during the handover process, data transmission is temporarily interrupted since the \ac{UE} cannot receive data from \ac{T-SBS} until \ac{UE} authentication and synchronization are completed (see Fig.~\ref{fig:3}).
Thus, the effective channel capacity $R_{\mathrm{eff}}^{(t)}(m^{(t-1:t)})$ at the $t$th time slot is expressed as a weighted sum of channel capacities during and after the handover process (i.e., $R^{(t)}(m^{(t)})(1-\mathbbm{1}_{\mathrm{ho}}(m^{(t-1:t)})$ and $R^{(t)}(m^{(t)})$):
\begin{IEEEeqnarray}{rCl}
    R_{\mathrm{eff}}^{(t)}\big(m^{(t-1:t)}\big) 
    &=&
    \frac{1}{\tau_s} \big(
    \tau_{\mathrm{ho}} R^{(t)}\big(m^{(t)}\big) (1-\mathbbm{1}_{\mathrm{ho}}(m^{(t-1:t)}))
    \ifthenelse{\equal{\main}{2}}
    {\notag\\ &&}{}
    + (\tau_s-\tau_{\mathrm{ho}}-\tau_{\mathrm{ce}}) R^{(t)}\big(m^{(t)}\big)  \big) \\
    &=& \frac{\tau_s - \tau_{\mathrm{ce}}}{\tau_s}(1-\mu\mathbbm{1}_{\mathrm{ho}}(m^{(t-1:t)})) R^{(t)}\big(m^{(t)}\big)\notag\\
    &&\IEEEeqnarraynumspace
\end{IEEEeqnarray}
where $\tau_{\mathrm{ho}}$ is the handover process time, and $\mu$ is the handover cost coefficient defined as
\begin{equation}
    \mu = \frac{\tau_{\mathrm{ho}}}{\tau_s - \tau_{\mathrm{ce}}}.
\end{equation}
Also, $\mathbbm{1}_{\mathrm{ho}}(m^{(t-1:t)})\in \{0,1\}$ is the binary handover indicator at the $t$th time slot defined as
\begin{align}
    \mathbbm{1}_{\mathrm{ho}}(m^{(t-1:t)}) \hspace{-0.2em} = \hspace{-0.2em}
    \begin{cases} 
    1 & \text{if $m^{(t-1)} \neq m^{(t)}$} \\
    0 & \text{otherwise}.
    \end{cases}
\end{align}
Note that in \ac{UDN} environments, \acp{UE} may undergo frequent handovers due to the dense deployment of \acp{SBS}, which consumes a considerable portion of the total transmission time and thus degrades throughput significantly.

\subsection{Conventional Mobility Management Techniques}\label{sec:II.C}
A major drawback of the 5G NR handover is that the signal quality deterioration and decision latency are considerable since handover is triggered only after the \ac{RSRP} drop has been reported.
This issue is even more pronounced in \ac{UDN} scenarios where handover occurs frequently due to the reduced cell coverage and increased number of \acp{SBS}.
To overcome the limitations of the reactive handover mechanism, proactive handover techniques that trigger handover before the signal quality degradation have been proposed~\cite{park2024mobility}.
Essence of this approach is to forecast the future \ac{RSRP} for all candidate \acp{SBS} using historical \ac{RSRP} measurements and then pick the \ac{SBS} that maximizes the throughput.
For the \ac{RSRP} prediction, \ac{RNN} architectures such as \ac{LSTM} and \ac{GRU} have been widely employed~\cite{shah2022multi, hu2023channel}.

While conventional proactive handover techniques are effective to some extent, they exclusively rely on \ac{RF} signals so that they fall short in capturing abrupt changes in the channel (e.g., blockages caused by dynamic obstacles).
Note that \ac{RF} signals capture gradual variations in signal quality, so it is very difficult to distinguish whether signal quality variations are transient (e.g., sudden blockages) or persistent (e.g., continuous path loss changes due to \ac{UE} movement) by merely checking the received signals~\cite{kim2024role}.
\ad{To address this issue, approaches that leverage sensing data (e.g., RGB images, radar signals, and LiDAR point clouds) obtained from various sensors (e.g., camera, radar, and LiDAR) have been proposed recently~\cite{charan2021vision, demirhan2022radar, liu2023vision,chaccour2024joint, feng2025networked}.}
These techniques focus primarily on identifying the \ac{LoS} component of the channel so that they suffer a degradation of the handover quality in \ac{NLoS} scenarios.

\vspace{-1em}
\subsection{Proactive Handover Decision Problem Formulation}\label{sec:II.D}
Effective mobility management must ensure seamless execution of handovers while avoiding unnecessary ones (e.g., ping-pong effects) caused by channel variations~\cite{liu2020proactive}.
To this end, we formulate a long-term proactive handover problem that not only responds to instantaneous degradations but also seeks to maximize channel capacity over extended periods.
Essence of the proactive handover problem $\mathscr{P}$ is to determine the \ac{SBS} indices $m^{(T:T+T_{\mathrm{p}})}$ for the time slots $T$ to $T+T_{\mathrm{p}}$, with an objective to maximize the cumulative channel capacity:
\begin{subequations} \label{opt}
\begin{IEEEeqnarray}{rCl}
    \mathscr{P}: & \hspace{-0.3em} \max_{m^{(T:T+T_{\mathrm{p}})}} & \sum_{t=T}^{T+T_{\mathrm{p}}} R_{\mathrm{eff}}^{(t)}(m^{(t-1:t)}) \IEEEeqnarraynumspace\\
    & \hspace{-0.3em} \textrm{s.t.} &  \hspace{-1.2em}R_{\mathrm{eff}}^{(t)}\big(m^{(t-1:t)}\big) \geq R_{\mathrm{min}}^{(t)} \hspace{0.5em} \forall t = T, \cdots, T+T_\mathrm{p} \IEEEeqnarraynumspace \label{constb}
\end{IEEEeqnarray}
\end{subequations}
where $R_{\mathrm{min}}^{(t)}$ is the minimum capacity requirement to ensure the \ac{QoS} of the \ac{UE} at the $t$th time slot.
The salient feature of $\mathscr{P}$ is that the objective function accounts for the cumulative channel capacity over $T_{\mathrm{p}}+1$ future time slots, which is clearly distinct from conventional approaches maximizing the instantaneous capacity (i.e., $R^{(T)}(m^{(T)})$) exclusively.
Using the cumulative channel capacity as a performance metric, we can reduce redundant handovers caused by instantaneous channel fluctuations.

To determine the optimal \ac{SBS} indices for future time slots, we need to know the future channel capacities (i.e., $\big\{ \{R^{(T+t)}(m^{(T+t)})\}_{t=0}^{T_{\mathrm{p}}} \mid m \in \mathcal{M} \}$).
Obviously, due to the causality issue, future channel capacities cannot be directly observed and need to be predicted instead.
Unfortunately, conventional approaches predicting future channel capacities based on temporal correlation (e.g., Kalman filter and \acp{RNN}) are not so effective, in particular for urban environments, due to the discontinuity of channel capacity (see Remark~\ref{remark1}).
\section{Large Multimodal Model-Based Environment-Aware Mobility Management}\label{sec:III}
The primary goal of \ac{LMM-EMM} is to proactively associate the \ac{UE} with proper \acp{SBS} that maximize channel capacity.
To this end, we leverage the multimodal reasoning capabilities of \ac{LMM}.
Specifically, we first extract the \ac{UE} mobility patterns and scattering geometry from the multimodal sensing data. 
Unlike conventional \ac{DL}-based approaches that only capture spatial correlations of the channel without any contextual information, \ac{LMM-EMM} exploits the road geometry and surrounding environment to infer the behavioral intention of the \ac{UE} (e.g., turning or accelerating) and perceive signal reflection patterns.
Next, using extracted features, we construct \ac{CCM}, a spatial mapping between the \ac{UE} position and the channel capacity, used for the estimation of the future channel capacity along the predicted \ac{UE} trajectory.
Furthermore, by analyzing RGB-D images with \ac{LMM}, we predict potential path blockages and then refine the channel capacity, thereby facilitating optimal \ac{SBS} selection even under high mobilities.

{\color{\blue}
A key feature of \ac{LMM-EMM} is that \ac{LMM} extracts environmental knowledge (e.g., the spatial layout of roads and buildings) from multimodal data and reuses this shared understanding across trajectory prediction, channel capacity estimation, and blockage prediction.
Unlike conventional \ac{DL} approaches that learn independent input-output mappings for each task without sharing physical knowledge, \ac{LMM-EMM} leverages a common environmental interpretation to produce mutually coherent predictions, thereby achieving more reliable mobility management performance.}
{\color{\blue}
The procedure of \ac{LMM-EMM} consists of four steps (see Fig.~\ref{fig:4}).
\begin{enumerate}
    \item \textbf{\ac{UE} trajectory prediction}: We predict the future \ac{UE} trajectory $\mathbf{p}_{\mathrm{ue}}^{(T:T+T_{\mathrm{p}})}$ using historical \ac{UE} positions $\mathbf{p}_{\mathrm{ue}}^{(T-T_\mathrm{w}:T-1)}$ and the \ac{BEV} map $\mathcal{I}_{\mathrm{bev}}$.
    \item \textbf{Static channel capacity estimation via \ac{CCM}}: By utilizing \ac{CCM}, we estimate the static channel capacities $\big\{ {R}_{\mathrm{ideal}}^{(T:
    T+T_{\mathrm{p}})}(m), {R}_{\mathrm{nlos}}^{(T:
    T+T_{\mathrm{p}})}(m)\mid m\in\mathcal{M}\big\}$ from the predicted \ac{UE} trajectories.
    \item \textbf{Blockage prediction and dynamic channel capacity refinement}: We predict path blockages from dynamic obstacles by tracking obstacle states from \ac{SBS}-view sensing images, and then derive the future achievable channel capacities $\big\{ {R}^{(T:T+T_{\mathrm{p}})}(m) \mid m\in\mathcal{M}\big\}$.
    \item \textbf{Proactive handover optimization}: We determine future \ac{SBS} indices $ m^{(T:T+T_{\mathrm{p}})}$ by solving $\mathscr{P}$ in~\eqref{opt} using \ac{DP}. 
    $\big\{ {R}^{(T:T+T_{\mathrm{p}})}(m) \mid m\in\mathcal{M}\big\}$ obtained in step 3 is used in this process.
\end{enumerate}
}

\subsection{LMM-Based UE Trajectory Prediction} \label{sec:IV.C}

\begin{figure}[t]
        \centering
        \ifthenelse{\equal{\main}{1}}
        {\newcommand{\mywidth}{0.5}}
        {\newcommand{\mywidth}{0.88}}
        \includegraphics[width=\mywidth\linewidth]{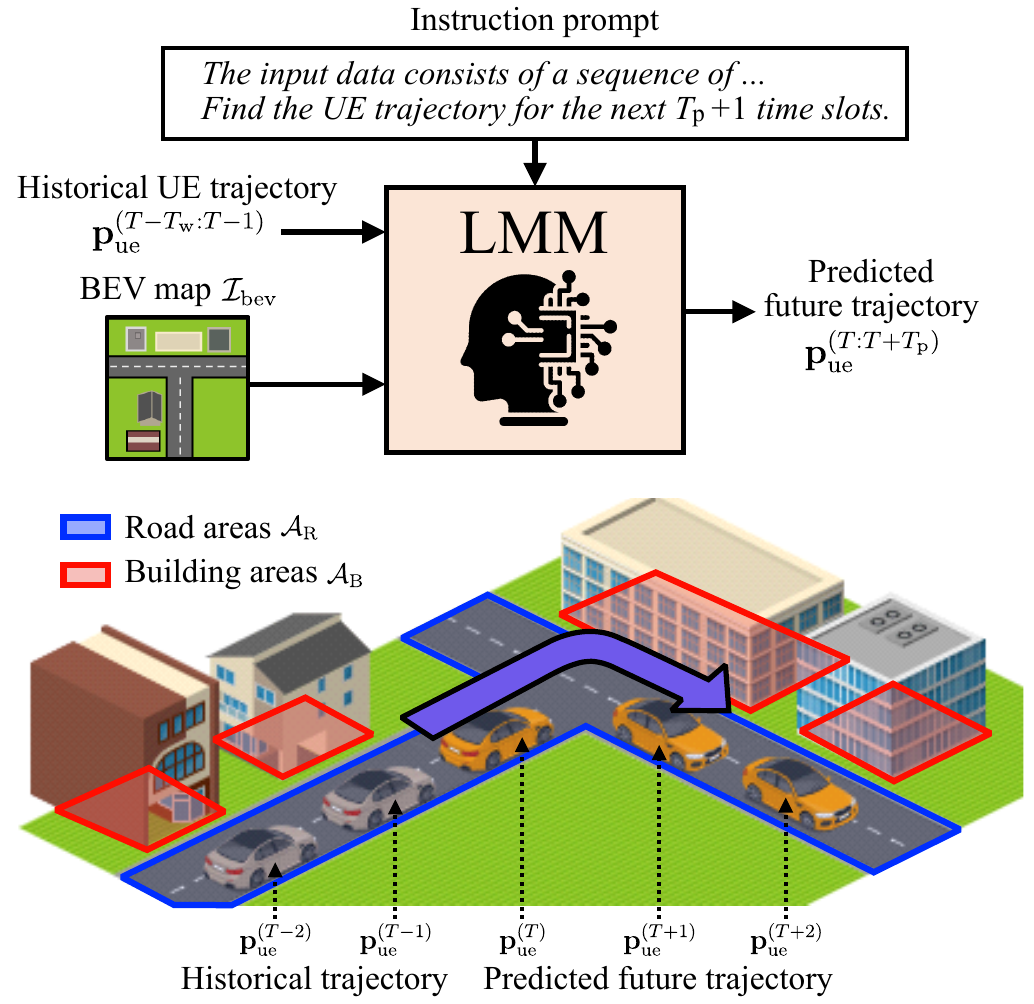}
        \caption{Illustration of LMM-based trajectory prediction using \ac{BEV} images.}
        \label{fig:5}
        \vspace{-1em} 
\end{figure}

In the \ac{UE} trajectory prediction step, we estimate the future \ac{UE} trajectory using historical \ac{UE} positions and the \ac{BEV} map:\footnote{\textcolor{\blue}{\ac{UE} position can be acquired from various positioning techniques leveraging \ac{GNSS}, sensing, and wireless measurements\cite{italiano2024tutorial,kim2024vision}.}}
\begin{IEEEeqnarray}{rCl} \label{traj}
    \mathbf{p}_{\mathrm{ue}}^{(T:T+T_{\mathrm{p}})} &=& f_{\mathrm{traj}} \big( \mathbf{p}_{\mathrm{ue}}^{(T-T_{\mathrm{w}}:T-1)}, \mathcal{I}_{\mathrm{bev}} \big) \IEEEeqnarraynumspace
\end{IEEEeqnarray}
where $f_\text{traj}$ is the \ac{UE} trajectory prediction function and $T_{\mathrm{w}}$ is the observation time window.
Also, $\mathcal{I}_{\mathrm{bev}}$ is the \ac{BEV} map image representing the static road topology, lane structures, and surrounding environments, which can be captured at the rooftop of a building or a satellite~\cite{zeng2024tutorial}.
It is important to note that the \ac{UE} movement is constrained by physical environments, such as road and building areas, which are denoted by $\mathcal{A}_{\mathrm{R}}$ and $\mathcal{A}_{\mathrm{B}}$, respectively~\cite{lan2024traj}.
For example, vehicles move along lanes with their trajectories confined within road boundaries (i.e., $\mathrm{Prob}(\mathbf{p}_{\mathrm{ue}}^{(t)}=\mathbf{p}\mid \mathbf{p}\notin\mathcal{A}_{\mathrm{R}}) = 0$), and pedestrians cannot traverse obstacles like buildings and walls (i.e., $\mathrm{Prob}(\mathbf{p}_{\mathrm{ue}}^{(t)}=\mathbf{p}\mid\mathbf{p} \in \mathcal{A}_{\mathrm{B}}) = 0$) (see Fig.~\ref{fig:5}).
In \ac{LMM-EMM}, we exploit \ac{BEV} images to incorporate $\mathcal{A}_{\mathrm{R}}$ and $\mathcal{A}_{\mathrm{B}}$ in \ac{UE} trajectory prediction.
This approach filters out impossible position predictions and also provides contextual cues for recognizing mobility patterns (e.g., turning at intersections)~\cite{keum2025deep}.
Specifically, we learn the conditional probability of the future \ac{UE} positions given past positions \ad{using the next token prediction capability of \ac{LMM} in an autoregressive manner:}
\begin{IEEEeqnarray}{rl}
    \mathrm{Prob} \big(&
    \mathbf{p}_{\mathrm{ue}}^{(T:T+T_{\mathrm{p}})}
    \mid \mathbf{p}_{\mathrm{ue}}^{(T-T_{\mathrm{w}}
    :T-1)}
    ,\mathcal{A}_{\mathrm{R}}, \mathcal{A}_{\mathrm{B}}\big) \notag \\
    &= \prod_{t=0}^{T_{\mathrm{p}}} 
    \mathrm{Prob} \big(\mathbf{p}_{\mathrm{ue}}^{(T+t)} \mid 
    \mathbf{p}_{\mathrm{ue}}^{(T-T_{\mathrm{w}}
    :T+t-1)}, 
    \mathcal{A}_{\mathrm{R}}, \mathcal{A}_{\mathrm{B}}\big). \IEEEeqnarraynumspace
\end{IEEEeqnarray}
\ad{Note that \ac{LMM} is trained with the autoregressive language modeling objective (i.e., next token prediction), where the model learns to predict the most likely subsequent token given all preceding tokens in a sequence~\cite{nie2023time}. 
This next token prediction mechanism of \ac{LMM} is particularly beneficial in \ac{LMM-EMM} since the next token prediction can be interpreted as predicting the future \ac{UE} position based on the historical \ac{UE} trajectory.}

\begin{figure}[t]
        \centering
        \ifthenelse{\equal{\main}{1}}
        {\newcommand{\mywidth}{0.5}}
        {\newcommand{\mywidth}{0.87}}        
        \includegraphics[width=\mywidth\linewidth]{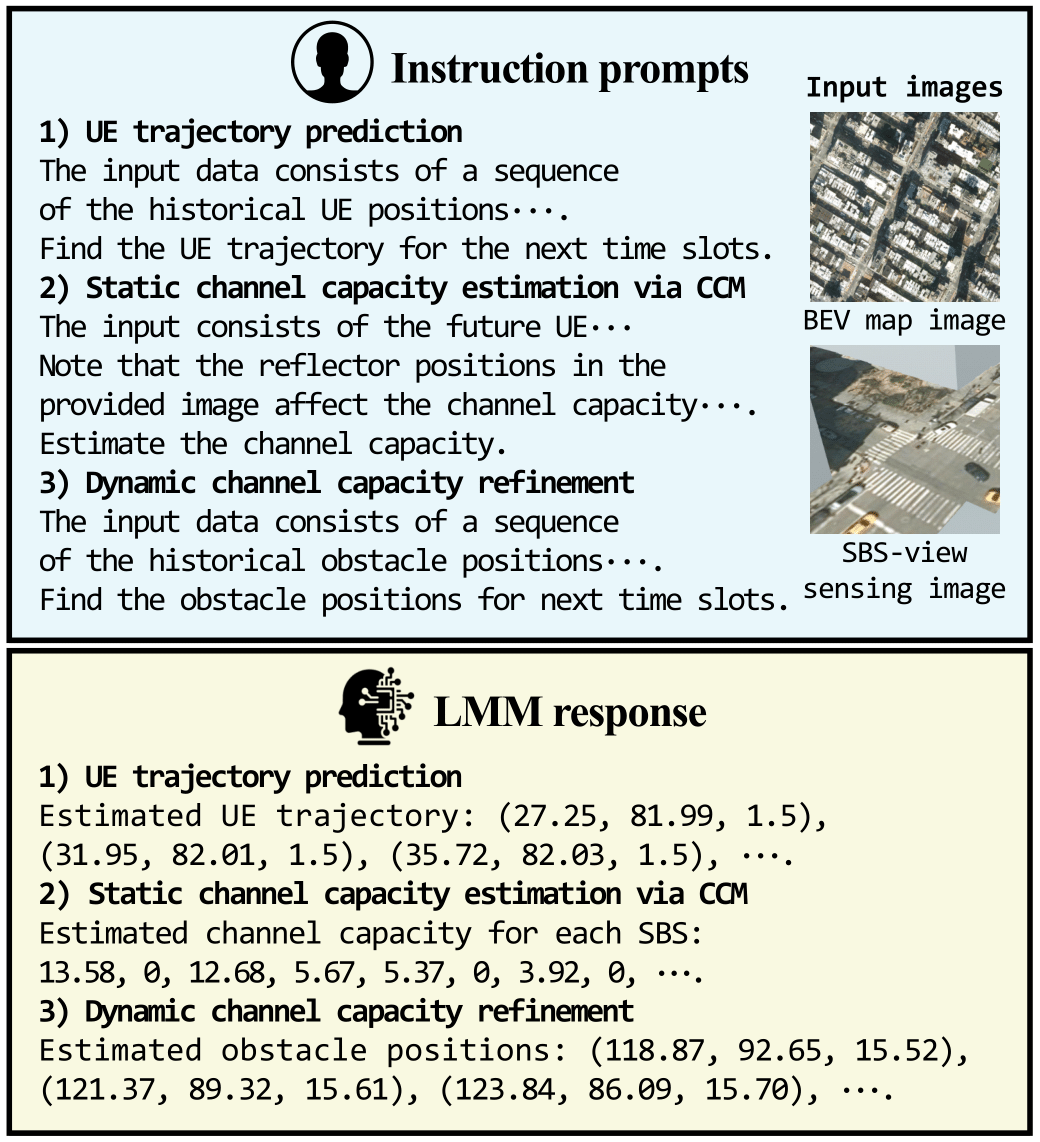}
        \caption{Examples of instruction prompt and \ac{LMM} response.}
        \vspace{-1.2em}
        \label{fig:6}
\end{figure}

\ad{An intriguing feature of \ac{LMM-EMM} is the instruction learning strategy that trains the model using natural language instructions specifying not only the input, task, and desired output, but also the task-relevant context.
In our case, the instruction prompts explicitly specify which environmental information should be emphasized in the multimodal data, allowing LMM to focus on the features most relevant to each task (see Fig.~\ref{fig:6}).}
For the \ac{UE} trajectory prediction task, the instruction prompts are structured as follows: 
\begin{itemize}
    \item \textbf{Input description}: \textit{``The input is a sequence of historical \ac{UE} positions from $(T-T_{\mathrm{w}})$th time slot to $(T-1)$th time slot $\mathbf{p}_{\mathrm{ue}}^{(T-T_{\mathrm{w}}:T-1)}$ and the \ac{BEV} map image $\mathcal{I}_{\mathrm{bev}}$."}
    \item \textbf{Environment description}: \textit{``The \ac{UE} movement is constrained by physical environments such as road boundaries $\mathcal{A}_{\mathrm{R}}$ and buildings $\mathcal{A}_{\mathrm{B}}$. Ensure the \ac{UE} trajectory is in feasible regions within the given environment."}
    \item \textbf{Task instruction}: \textit{``Find the \ac{UE} trajectory for the next $T_{\mathrm{p}}+1$ time slots $\mathbf{p}_{\mathrm{ue}}^{(T:T+T_{\mathrm{p}})}$."}
\end{itemize}
Given the instruction prompts, the \ac{LMM} generates the response prompt as \textit{``The predicted \ac{UE} trajectory for the next $T_\mathrm{p}+1$ time slots is $\mathbf{p}_{\mathrm{ue}}^{(T:T+T_{\mathrm{p}})}$."}

\subsection{LMM-Based Static Channel Capacity Estimation via CCM} \label{sec:IV.D}
In the channel capacity estimation step, we learn the \ac{CCM} $f_{\mathrm{ccm}}$ in \eqref{composition}, which maps the \ac{UE} and \ac{SBS} positions $(\mathbf{p}_{\mathrm{ue}}^{(t)},\mathbf{p}_{\mathrm{sbs},m})$ to the static channel capacities $(R_{\mathrm{ideal}}^{(t)}(m),R_{\mathrm{nlos}}^{(t)}(m))$, parametrized by reflector information $\mathcal{E}$.
Two main challenges in learning $f_{\mathrm{ccm}}$ are: 1) the discontinuous variation of static channel capacities with \ac{UE} mobility (see Remark~\ref{remark1}), and 2) the difficulty of directly acquiring reflector information $\mathcal{E}$, as accurately modeling and extracting $\mathcal{E}$ in real environments is highly complex.
To address this issue, we learn a surrogate function $\tilde{f}_{\mathrm{ccm}}$ that takes the \ac{BEV} map image, which implicitly encodes $\mathcal{E}$:
\begin{IEEEeqnarray}{rCl}
\big ({R}_{\mathrm{ideal}}^{(t)}(m), {R}_{\mathrm{nlos}}^{(t)}(m) \big) &=& \tilde{f}_{\mathrm{ccm}}\big(\mathbf{p}_{\mathrm{ue}}^{(t)}, \mathbf{p}_{\mathrm{sbs},m}, \mathcal{I}_{\mathrm{bev}}\big). \IEEEeqnarraynumspace
\end{IEEEeqnarray}
Furthermore, we exploit the multimodal reasoning capability of \acp{LMM} to interpret reflector information from the \ac{BEV} map and to pinpoint locations where channel capacity discontinuities may occur.
In doing so, \ac{LMM-EMM} integrates environmental context into the surrogate model, facilitating reliable approximation of $f_{\mathrm{ccm}}$ and precise estimation of static channel capacities at arbitrary \ac{UE} and \ac{SBS} locations.
The instruction prompts for the \ac{LMM}-based channel capacity estimation are structured as follows:
\begin{itemize}
    \item \textbf{Input description}: \textit{``The input is the future \ac{UE} position $\mathbf{p}_{\mathrm{ue}}^{(t)}$ and the \ac{BEV} map $\mathcal{I}_{\mathrm{bev}}$."}
    \item \textbf{Environment description}: \textit{``Note that the reflector positions in the provided image affect the channel capacity by reflecting or blocking the propagation paths."}
    \item \textbf{Task instruction}: \textit{``Estimate the static channel capacities $ \{({R}_{\mathrm{ideal}}^{(t)}(m), {R}_{\mathrm{nlos}}^{(t)}(m)) \mid m \in \mathcal{M} \}$ from the \ac{UE} position."}
\end{itemize}
Given the instruction prompts, we obtain the estimated channel capacities from the \ac{LMM} response: \textit{``The estimated static channel capacities are $\big\{({R}_{\mathrm{ideal}}^{(t)}(m), {R}_{\mathrm{nlos}}^{(t)}(m)) \mid m \in \mathcal{M} \big\}$."}

\subsection{LMM-Based Blockage Prediction and Dynamic Channel Capacity Refinement} \label{sec:IV.E}
In the blockage prediction and dynamic channel capacity refinement step, we first estimate the \ac{LoS} indicators in~\eqref{clos} $\big\{{c}_{\mathrm{los}}^{(T:T+T_{\mathrm{p}})}(m) \mid m \in \mathcal{M} \big\}$ from \ac{SBS}-view sensing images.
Using the estimated \ac{LoS} indicators, together with the static channel capacity estimates $\big\{ R_{\mathrm{ideal}}^{(T:T+T
_{\mathrm{p}})}(m),R_{\mathrm{nlos}}^{(T:T+T
_{\mathrm{p}})}(m) \mid m \in \mathcal{M} \big\}$, we then compute the achievable channel capacity $\big \{{R}^{(T:T+T_{\mathrm{p}})}(m) \mid m \in \mathcal{M} \big\}$ according to \eqref{dynamic_capacity}.

\begin{figure}[t]
        \centering
        \ifthenelse{\equal{\main}{1}}
        {\newcommand{\mywidth}{0.4}}
        {\newcommand{\mywidth}{0.75}}
        \includegraphics[width=\mywidth\linewidth]{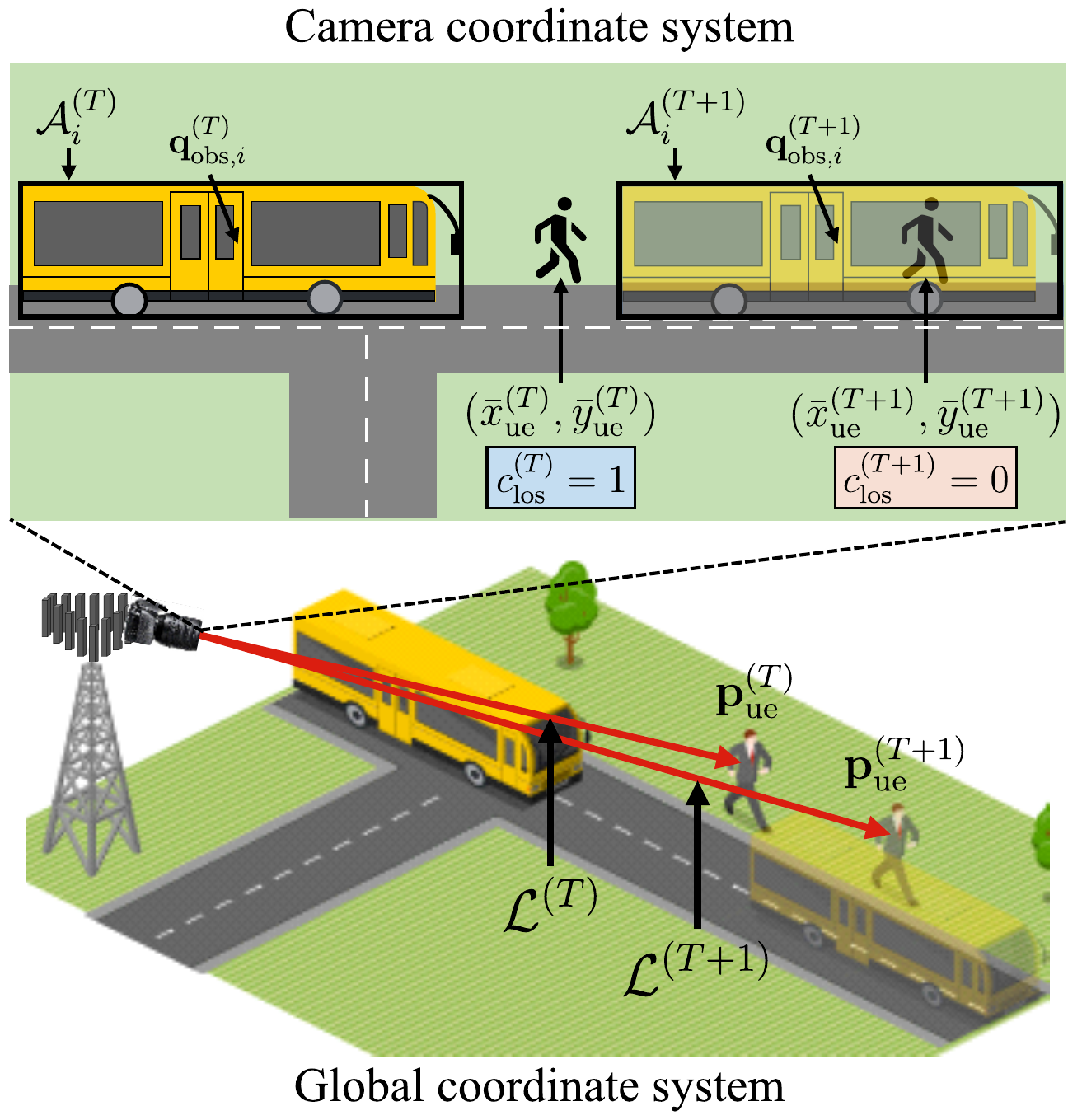}
        \caption{Illustration of LMM-based dynamic blockage prediction with \ac{SBS}-view sensing images.}
        \vspace{-1em}
        \label{fig:7}
\end{figure}

Specifically, let $\mathcal{N}_{\mathrm{obs},m}=\{1,\dotsc,N_{\mathrm{obs},m}\}$ be the set of objects near the \ac{SBS} $m$. 
For each object $i\in\mathcal{N}_{\mathrm{obs},m}$, we denote the center position of the object $i$ in the camera coordinate system as 
\begin{equation}
\mathbf{q}_{\mathrm{obs},i}^{(t)}=\big[x_{\mathrm{obs}, i}^{(t)} \,\, y_{\mathrm{obs}, i}^{(t)} \,\, d_{\mathrm{obs}, i}^{(t)}\big]   
\end{equation}
where $x_{\mathrm{obs}, i}^{(t)}$, $y_{\mathrm{obs}, i}^{(t)}$, and $d_{\mathrm{obs}, i}^{(t)}$ are x-coordinate, y-coordinate, and depth of the object centroid, respectively.
Then, the 2D rectangular region $\mathcal{A}_i^{(t)}\subset \mathbb{R}^2$ in the image occupied by the object $i$ at time $t$ is given by
\begin{IEEEeqnarray}{rl}
\mathcal{A}_i^{(t)} =& \Big[x_{\mathrm{obs}, i}^{(t)}-\frac{1}{2}w_{\mathrm{obs}, i}^{(t)}, x_{\mathrm{obs}, i}^{(t)}+\frac{1}{2}w_{\mathrm{obs}, i}^{(t)}\Big] \notag\\
&\times \Big[y_{\mathrm{obs}, i}^{(t)}-\frac{1}{2}h_{\mathrm{obs}, i}^{(t)}, y_{\mathrm{obs}, i}^{(t)}+\frac{1}{2}h_{\mathrm{obs}, i}^{(t)}\Big]
\end{IEEEeqnarray}
where $w_{\mathrm{obs}, i}^{(t)}$ and $h_{\mathrm{obs}, i}^{(t)}$ are the width and height of the bounding box of the object $i$ in the camera coordinate system, respectively.
Using this, $c_{\mathrm{los}}^{(t)}(m)$ is defined as
\begin{equation} 
\setlength{\nulldelimiterspace}{0pt} 
c_{\mathrm{los}}^{(t)}(m) \hspace{-0.2em}=\hspace{-0.3em}\left\{\begin{IEEEeqnarraybox}[\relax][c]{
 l's} 
0& \hspace{-1em} if \hspace{-0.4em}
$(\bar{x}_{\mathrm{ue}}^{(t)}, \bar{y}_{\mathrm{ue}}^{(t)}) \hspace{-0.2em}\in\hspace{-0.2em} \mathcal{A}_i^{(t)}$ \hspace{-0.6em} and \hspace{-0.4em}
$d_{\mathrm{obs}, i}^{(t)} \hspace{-0.2em}<\hspace{-0.2em} d_{\mathrm{ue}}^{(t)}$ \hspace{-0.5em}
$\exists i \hspace{-0.1em}\in \hspace{-0.2em}\mathcal{N}_{\mathrm{obs},m}$ \IEEEeqnarraynumspace\\
1& \hspace{-1em} otherwise 
\end{IEEEeqnarraybox}\right.
\end{equation}
where $d_{\mathrm{ue}}^{(t)}$$\,=$\,$\|\mathbf{p}_{\mathrm{ue}}^{(t)} - \mathbf{p}_{\mathrm{sbs},m}^{(t)}\|_2$ and $(\bar{x}_{\mathrm{ue}}^{(t)}, \bar{y}_{\mathrm{ue}}^{(t)})$ is the x- and y-coordinates of \ac{UE} in the camera coordinate system at time slot $t$ (see Fig.~\ref{fig:7}).
Note that $(\bar{x}_{\mathrm{ue}}^{(t)}, \bar{y}_{\mathrm{ue}}^{(t)})$ can be obtained from the \ac{GCS} \ac{UE} position $\mathbf{p}_{\mathrm{ue}}^{(t)}$ through rotation as $(\bar{x}_{\mathrm{ue}}^{(t)}, \bar{y}_{\mathrm{ue}}^{(t)}) = \big(\frac{x_r^{(t)}}{z_r^{(t)}}, \frac{y_r^{(t)}}{z_r^{(t)}}\big)$ where
\begin{IEEEeqnarray}{rCl}
    \big[x_r^{(t)} \,\, y_r^{(t)} \,\, z_r^{(t)}\big]^{\mathrm{T}}
    &=&
    \mathbf{R}\big(\mathbf{p}_{\mathrm{ue}}^{(t)} - \mathbf{p}_{\mathrm{sbs},m}\big).
\end{IEEEeqnarray}
Here, $\mathbf{R}$ is the camera rotation matrix given by~\cite{ahn2024sensing}
\begin{equation}
    \mathbf{R} = \begin{bmatrix}
        \sin\theta & -\cos\theta & 0 \\
        -\sin\phi \cos\theta & -\sin\phi \sin\theta & \cos\phi \\
        \cos\phi \cos\theta & \cos\phi \sin\theta & \sin\phi
    \end{bmatrix} 
\end{equation}
where $\theta$ and $\phi$ denote the yaw and pitch angles of the camera.

To determine ${c}_{\mathrm{los}}^{(t)}(m)$, the \ac{SBS} $m$ needs to know the future positions of objects $\mathbf{q}_{\mathrm{obs},i}^{(T:T+T_{\mathrm{p}})}$ for $i\in\mathcal{N}_{\mathrm{obs},m}$. 
Analogous to \ac{UE} trajectory prediction, these positions are inferred from historical positions $\mathbf{q}_{\mathrm{obs},i}^{(T-T_{\mathrm{w}}:T-1)}$.
Unlike \ac{UE} position data, which can be reported from \ac{UE} via uplink channel, historical object positions are difficult to obtain since no communication links exist between the \ac{SBS} and the objects~\cite{kim2025large2}.
As a remedy, we utilize \ac{OD} techniques to accurately extract $ \mathbf{q}_{\mathrm{obs},i}^{(T-T_{\mathrm{w}}:T-1)}$ from \ac{SBS}-view RGB-D images $\mathcal{I}_{\mathrm{sbs},m}$, and then use \ac{LMM} to predict the future object trajectories:
\begin{IEEEeqnarray}{rCl}
    \mathbf{q}_{\mathrm{obs},i}^{(T:T+T_{\mathrm{p}})}
    &=&f_{\mathrm{obs}} \big( \mathbf{q}_{\mathrm{obs},i}^{(T-T_{\mathrm{w}}:T-1)}, \mathcal{I}_{\mathrm{sbs},m} \big)
    \label{traj2d} \IEEEeqnarraynumspace
\end{IEEEeqnarray}
where $f_{\mathrm{obs}}$ is the obstacle trajectory prediction function.
The instruction prompts for the \ac{LMM}-based obstacle position prediction are designed as follows: 
\begin{itemize}
    \item \textbf{Input description}: \textit{``The input is a sequence of the historical obstacle positions in the camera coordinate system $\mathbf{q}_{\mathrm{obs},i}^{(T-T_{\mathrm{w}}:T-1)}$ and the \ac{SBS}-view sensing image $\mathcal{I}_{\mathrm{sbs}}$."}
    \item \textbf{Environment description}: \textit{``In the camera coordinate system, obstacle movement is constrained by environmental elements including road boundaries for vehicles and immobile structures such as buildings that define non-navigable regions observed in $\mathcal{I}_{\mathrm{sbs}}$."}
    \item \textbf{Task instruction}: \textit{``Find the positions of dynamic obstacles for the next $(T_{\mathrm{p}}+1)$ time slots $\mathbf{q}_{\mathrm{obs},i}^{(T:T+T_{\mathrm{p}})}$."}
\end{itemize}
The \ac{LMM} response prompts generated from the instruction prompts are \textit{``The positions of dynamic obstacles for the next $T_{\mathrm{p}}+1$ time slots are $\mathbf{q}_{\mathrm{obs},i}^{(T:T+T_{\mathrm{p}})}$."}

Once we obtain the predicted static channel capacities and \ac{LoS} indicators $\big\{ R_{\mathrm{ideal}}^{(T:T+T
_{\mathrm{p}})}(m),R_{\mathrm{nlos}}^{(T:T+T
_{\mathrm{p}})}(m), c_{\mathrm{los}}^{(T:T+T_{\mathrm{p}})}(m) \mid m \in \mathcal{M} \big\},$ for all \acp{SBS}, we acquire the achievable channel capacity $\big\{{R}^{(T:T+T_{\mathrm{p}})}(m) \mid m \in \mathcal{M} \big\}$ using~\eqref{dynamic_capacity}.

\vspace{-1em}
\subsection{Proactive Handover Optimization} \label{sec:IV.F}

In the proactive handover optimization step, we find the optimal \ac{SBS} indices $\hat{m}^{(T:T+T_{\mathrm{p}})}$ in future time slots by solving the optimization problem $\mathscr{P}$ in \eqref{opt} based on the estimated channel capacities $\big\{{R}^{(T:T+T_{\mathrm{p}})}(m) \mid m \in \mathcal{M} \big\}$.
One intuitive yet naive approach to solve $\mathscr{P}$ is an exhaustive search, which evaluates all possible sequences of \acp{SBS}.
Although exhaustive search guarantees the optimal solution, it becomes computationally prohibitive in \ac{UDN} because the number of candidates grows exponentially as $M^{T_{\mathrm{p}}+1}$.

To obtain a tractable solution of $\mathscr{P}$, we adopt a \ac{DP}-based approach, which solves complex problems by decomposing them into a series of simpler subproblems~\cite{bellman1966dynamic}.
Owing to the cumulative structure of the capacity, the optimal solution at each time slot can be recursively derived from the solutions of previous time slots.
In fact, the computational complexity of the proposed \ac{DP}-based method is $T_{\mathrm{p}}M^2$, which is significantly lower than that of the exhaustive search.
Specifically, let $g({T+t}, m^{(T+t)})$ be the maximum cumulative capacity when the \ac{UE} is connected to the \ac{SBS} $m^{(T+t)}$ at time slot $T+t$:
\begin{IEEEeqnarray}{rl}
    g\big({T+t}, m^{(T+t)}\big)
    = \hspace{-1.02em} \max_{m^{(T:T+t-1)}}  \, \Bigg[ & \sum_{t'=0}^{t} R_{\mathrm{eff}}^{(T+t')}\big(m^{(T+t'-1:T+t')}\big) \notag \\
    &\times \mathbbm{1}_{\mathrm{min}}\big(m^{(T+t'-1:T+t')}\big) \Bigg] \IEEEeqnarraynumspace
\end{IEEEeqnarray}
where $\mathbbm{1}_{\mathrm{min}}(m^{(t-1:t)})$ is the capacity compliance indicator defined as
\begin{IEEEeqnarray}{rl}
    \mathbbm{1}_{\mathrm{min}}(m^{(t-1:t)}) = 
    \begin{cases} 
    1 & \text{if $R_{\mathrm{eff}}^{(t)}\big(m^{(t-1:t)}\big) \geq R_{\mathrm{min}}^{(t)}$} \\
    -\infty & \text{if $R_{\mathrm{eff}}^{(t)}\big(m^{(t-1:t)}\big) < R_{\mathrm{min}}^{(t)}$}.
    \end{cases} \IEEEeqnarraynumspace
\end{IEEEeqnarray}
Note that $g({T+t}, m^{(T+t)})$ can be decomposed as
\begin{IEEEeqnarray}{rl} \label{decompose}
    &g({T+t}, m^{(T+t)}) = \hspace{-0.2em} \max_{m^{(T+t-1)}} \Big[ g\big(T+t-1, m^{(T+t-1)}\big) \notag \\
    & \hspace{1em}+ R_{\mathrm{eff}}^{(T+t)}\big(m^{(T+t-1:T+t)}\big) \mathbbm{1}_{\mathrm{min}}\big(m^{(T+t-1:T+t)}\big) \Big]. \IEEEeqnarraynumspace
\end{IEEEeqnarray}
Based on this recurrence formula, starting from initial $g(T, m^{(T)}) = R_{\mathrm{eff}}^{(T)}(m^{(T-1:T)}) \mathbbm{1}_{\mathrm{min}}(m^{(T-1:T)})$, we sequentially compute
$\big\{\{ g(T+t, m^{(T+t)}) \}_{t=1}^{T_\mathrm{p}} \mid m \in \mathcal{M} \big\}$, which are subsequently stored in the \ac{DP} table.
After that, $\hat{m}^{(T:T+T_\mathrm{p})}$ are retrieved in reverse chronological order from the computed \ac{DP} table.
First, $\hat{m}^{(T+T_{\mathrm{p}})}$ is obtained as
\begin{IEEEeqnarray}{rCl}
    \hat{m}^{(T+T_\mathrm{p})} &=& \argmax_{m^{(T+T_\mathrm{p})}}\,\, g\big({T+T_\mathrm{p}}, m^{(T+T_\mathrm{p})}\big).
\end{IEEEeqnarray}
Then, $\hat{m}^{(t)}$ are determined recursively for $t=T_{\mathrm{p}}-1, \cdots, 0$:
\begin{IEEEeqnarray}{rCl}
    \hat{m}^{(T+t)} &=& \argmax_{m^{(T+t)}} \,\,\Big[ g(T+t, m^{(T+t)}) \notag \\
    && \hspace{-1.5em}+ R_{\mathrm{eff}}^{(T+t+1)}(m^{(T+t:T+t+1)}) \mathbbm{1}_{\mathrm{min}}(m^{(T+t:T+t+1)}) \Big]. \IEEEeqnarraynumspace
\end{IEEEeqnarray}
Finally, the \acp{SBS} execute handovers according to the determined indices $\hat{m}^{(T:T+T_\mathrm{p})}$.

{\color{\blue}
\section{Practical Issues}\label{sec:IV}
In this section, we discuss several practical issues related to the implementation of \ac{LMM-EMM}, focusing on end-to-end latency and challenges associated with its deployment.}

{\color{\blue}
\subsection{End-to-end Latency}
The key requirement for reliable handover decision-making is that the total end-to-end latency of \ac{LMM-EMM} remain shorter than the timescale over which the channel varies sufficiently to affect the ergodic channel capacity.
As shown in Equation~\eqref{beamcohtime}, the ergodic channel capacity is determined by slow-varying channel parameters, such as angles and path losses, whose coherence time $T_{\mathrm{c}}$ (i.e., beam coherence time) is given by~\cite{va2016impact}
\begin{equation}
    T_\mathrm{c} = \frac{D}{v\cos \theta} \cos^{-1} (\psi^2 \log \zeta + 1)
    \label{beamcohtime}
\end{equation}
where $D$ is the communication distance, $\theta$ is the beam angle, $v$ is the speed of the \ac{UE}, $\psi$ is the beamwidth, and $\zeta \in [0,1]$ is the threshold ratio of the received power to its peak value.
For example, when a UE moves at $25\,\text{km/h}$ with $D=70\,\text{m}$, $\theta=60^\circ$, and $\zeta=0.8$, and the SBS is equipped with $N=32$ antennas such that $\psi \approx \frac{4}{N} = 0.125\,\mathrm{rad}$, then the beam coherence time is $T_\mathrm{c} \approx 840\,\text{ms}$.

In our implementation, the latency of each LMM-EMM module on an NVIDIA L40S GPU is summarized as follows:
\begin{itemize}
    \item \textbf{UE trajectory prediction}: $T_1=215\,$ms ($20\,$ms for UE location report and $195\,$ms for LMM inference)
    \item \textbf{Static channel capacity estimation}: $T_2=195\,$ms ($195\,$ms for LMM inference)
    \item \textbf{Blockage prediction}: $T_3=220\,$ms ($25\,$ms for multimodal image capture and processing, and $195\,$ms for LMM inference)
    \item \textbf{Channel capacity refinement, report, and handover}: $T_4=55\,$ms ($25\,$ms for channel capacity refinement and report, and $30\,$ms for DP-based handover decision)
\end{itemize}
Since blockage prediction can be performed in parallel with UE trajectory prediction and static channel capacity estimation, as it does not depend on others' outputs, these three modules are completed within $\max(T_1 + T_2, T_3) = 410\,$ms.
Therefore, the total end-to-end latency of \ac{LMM-EMM} is $T_{\mathrm{tot}}=\max(T_1 + T_2, T_3) + T_4 = 465\,$ms, which is substantially shorter than the beam coherence time $T_{\mathrm{c}}\approx 840\,$ms (see Fig.~\ref{beamcoh}).
Moreover, this latency is expected to decrease further as GPU hardware continues to evolve, leaving considerable room for further optimization in future implementations.

\begin{figure}[t]
    \centering
    \includegraphics[width=\linewidth]{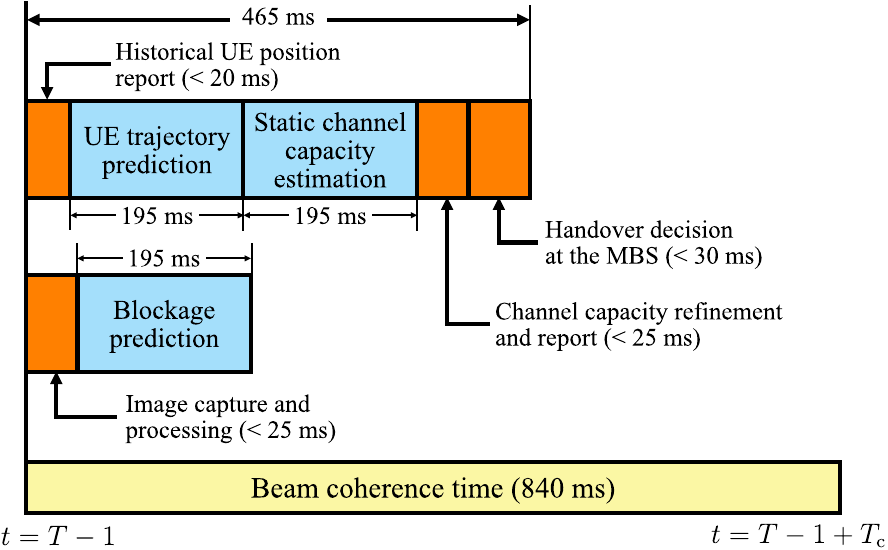}
    \caption{{\color{\blue}End-to-end latency vs. beam coherence time.}}
    \label{beamcoh}
    \vspace{-1em}
\end{figure}

\vspace{-1em}
\subsection{Deployment Issues}
\subsubsection{Noisy Sensing}
In real-world settings, sensing images are often degraded by environmental factors such as rain, snow, fog, or dust.
To demonstrate the robustness of the proposed LMM-EMM framework to imperfections in sensing data, we evaluate the handover performance of LMM-EMM under noisy sensing conditions caused by environmental factors.
Under this setting, LMM-EMM achieves the average channel capacity of $4.511\,$bps/Hz, which corresponds to only a 4\% reduction from the noise-free case.
Furthermore, when the denoising and restoration techniques for adverse visual conditions are applied, \ac{LMM-EMM} achieves the average channel capacity of $4.724\,$bps/Hz (see Fig.~\ref{noisy})~\cite{patil2023multi}.
This corresponds to only a $1\%$ decrease compared to $4.770\,$bps/Hz in the noise-free scenario (see Fig.~\ref{fig:13}).

\subsubsection{FoV and Resolution Issues}
One practical issue of \ac{LMM-EMM} is that the \ac{SBS}-view camera has a limited \ac{FoV} and resolution, which may restrict its ability to capture the entire scene.
While it is true that a single camera has a limited \ac{FoV}, the entire coverage area can be monitored by deploying multiple cameras, each covering a different sectorized region, similar to antenna sectorization.
For example, three RGB-D cameras, each with a \ac{FoV} of $120^\circ$, can cover the entire area surrounding the SBS.

Regarding the image resolution issue, objects that contribute to LoS blockage (e.g., car, bus, and truck) can still be effectively identified even in low-resolution images because they are typically large enough to occupy multiple pixels in the captured images.
Furthermore, in our work, we crop and resize the image region around the predicted UE position to magnify relevant objects and improve detection performance (see Fig.~\ref{crop}).


\begin{figure}[t]
    \centering
    \begin{subfigure}{\linewidth}
        \centering
        \includegraphics[width=0.9\linewidth]{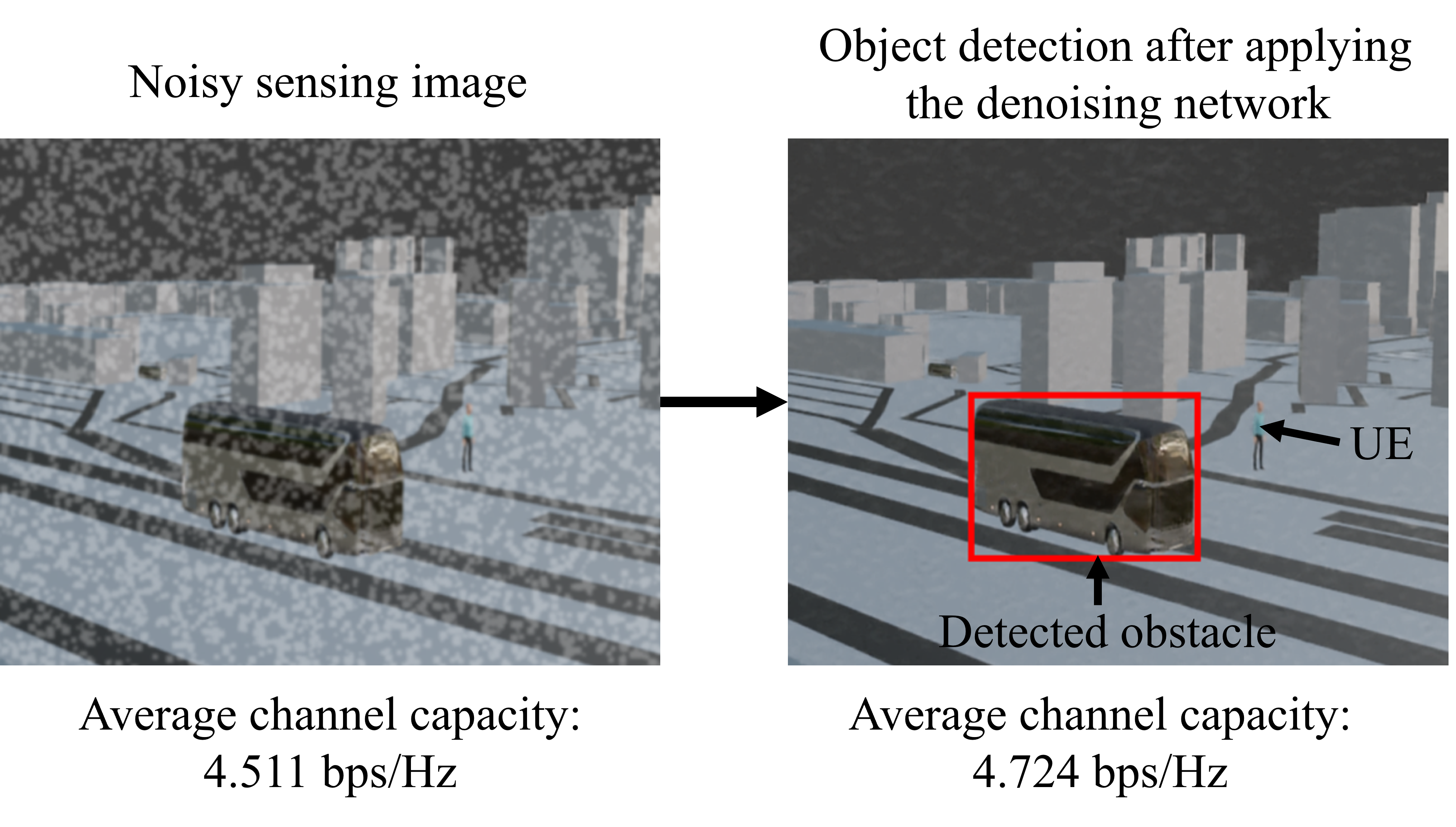}
        \caption{Illustration of the noisy sensing image and object detection after applying the denoising network.}
        \label{noisy}
    \end{subfigure}
    
    \vspace{0.5em}
    
    \begin{subfigure}{\linewidth}
        \centering
        \includegraphics[width=0.9\linewidth]{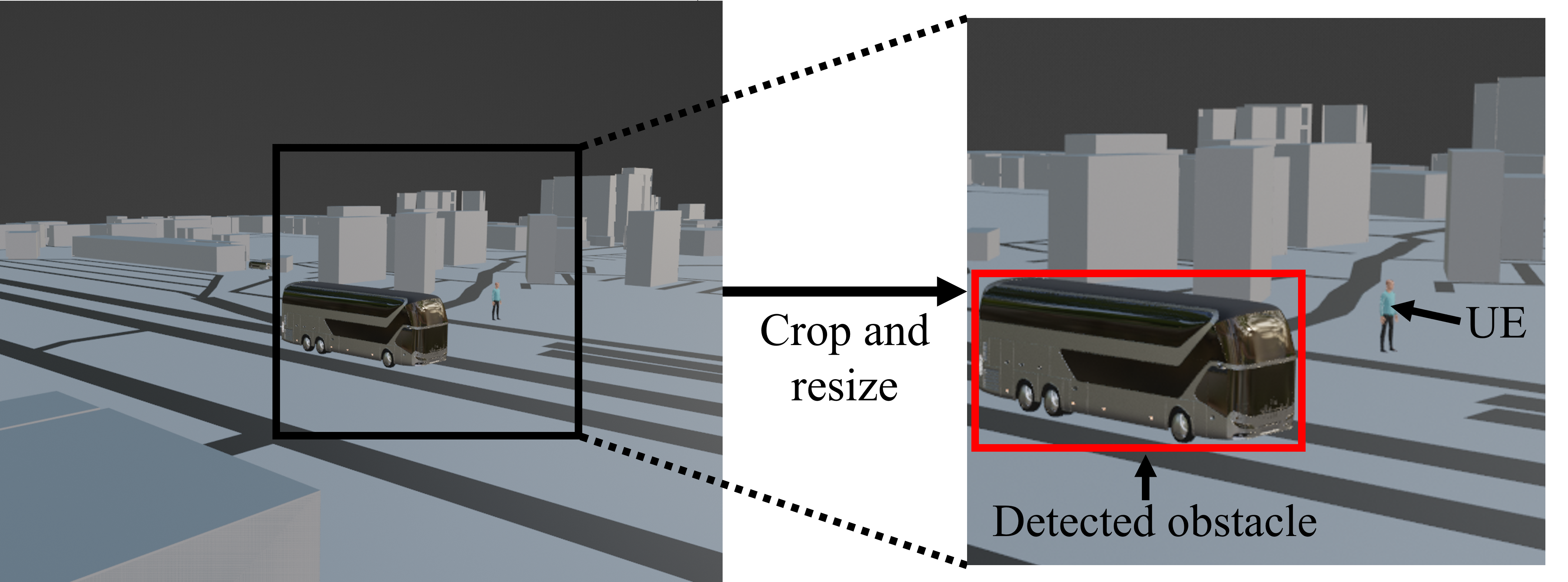}
        \caption{Illustration of the cropped image region around the predicted UE position.}
        \label{crop}
    \end{subfigure}
    
    \caption{{\color{\blue}Examples of sensing images used for blockage prediction under noisy and low-resolution scenarios.}}
    \vspace{-1em}
    \label{reflector}
\end{figure}

\subsubsection{Availability of Cameras and BEV Maps}
One might be concerned that LMM-EMM can operate only when RGB-D cameras are installed and \ac{BEV} maps are available at the \acp{SBS}.
\adtwo{However, with \ac{ISAC} emerging as a key component of upcoming 6G, \acp{SBS} are expected to be equipped with diverse sensing modalities including RGB-D cameras.
Moreover, \ac{BEV} maps can be readily obtained from publicly available sources such as OpenStreetMap~\cite{haklay2008openstreetmap} and Google Maps.}

\adtwo{If the serving \ac{SBS} cannot collect the sensing data due to the absence of sensors, it can request a neighboring \ac{SBS} that can observe the UE and surrounding obstacles to estimate the blockage state.
Specifically, the serving \ac{SBS} transmits the estimated UE trajectory to the neighboring \ac{SBS}. 
The neighboring \ac{SBS} then constructs the LoS line segment between the serving \ac{SBS} and the UE, and uses its own RGB-D camera to detect obstacles and represent them as 3D boxes.
By checking whether the LoS line segment intersects the 3D boxes, the blockage state can be determined~\cite{kim2025large2}.
If no such neighbor exists, the blockage state can be inferred from past channel measurements (e.g., \ac{SINR}), and the inferred past state can be used as a proxy for the future LoS state.
Such temporal consistency is reasonable because obstacles that cause blockage (e.g., buses, vehicles) typically have non-negligible volume, so the blockage state does not change significantly over a short time interval.}

\subsubsection{{\color{\bluetwo}Robustness to Reflector Material Mismatch}}
\adtwo{A possible issue in \ac{LMM-EMM} is that incorrect knowledge about the material properties may introduce errors in the channel capacity estimation.
However, since material properties generally do not change significantly over time, the associated errors can be controlled via fine-tuning with real datasets.
Even if the reflector material is incorrectly characterized, \ac{LMM-EMM} can still maintain reasonable performance because the propagation geometry (e.g., LoS paths, reflection angles) remains unchanged despite the material mismatch.
To validate this, we conduct an ablation study in which the reflector material in the validation dataset is changed from glass to concrete.
We observe a performance degradation of 5.2\% compared with the original case, which demonstrates that the proposed technique still maintains reasonable performance under incorrect knowledge of reflector materials (see Fig.~\ref{reflector_graph}).}

\subsubsection{Synchronization} 
\adtwo{According to the IEEE 1588-v2 protocol, inter-cell time synchronization errors within $0.1\,\mu$s may exist~\cite{SMPTE_EG2111_1_2021}.
To evaluate the impact of these synchronization errors, we measure the average channel capacity under such timing offsets.
We observe that the average channel capacity remains $4.770\,\mathrm{bps/Hz}$, identical to the perfectly synchronized case of $4.770\,\mathrm{bps/Hz}$, indicating no performance degradation.
This is because the synchronization error is sufficiently small that the \ac{UE} displacement during this interval is negligible, and thus the corresponding channel can be considered as unchanged.}


\begin{figure*}[h]
\centering
\begin{minipage}[t]{0.315\textwidth}
    \centering
    \includegraphics[width=\textwidth]{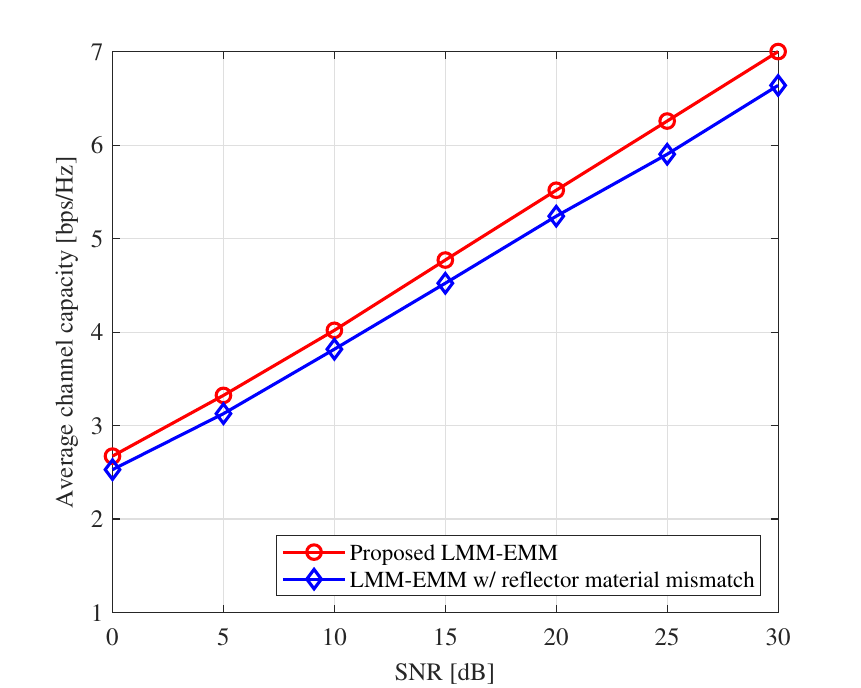}
    \caption{{\color{\bluetwo}Average channel capacity of LMM-EMM with reflector material mismatch.}}
    \label{reflector_graph}
\end{minipage}
\hfill
\begin{minipage}[t]{0.315\textwidth}
    \centering
    \includegraphics[width=\textwidth]{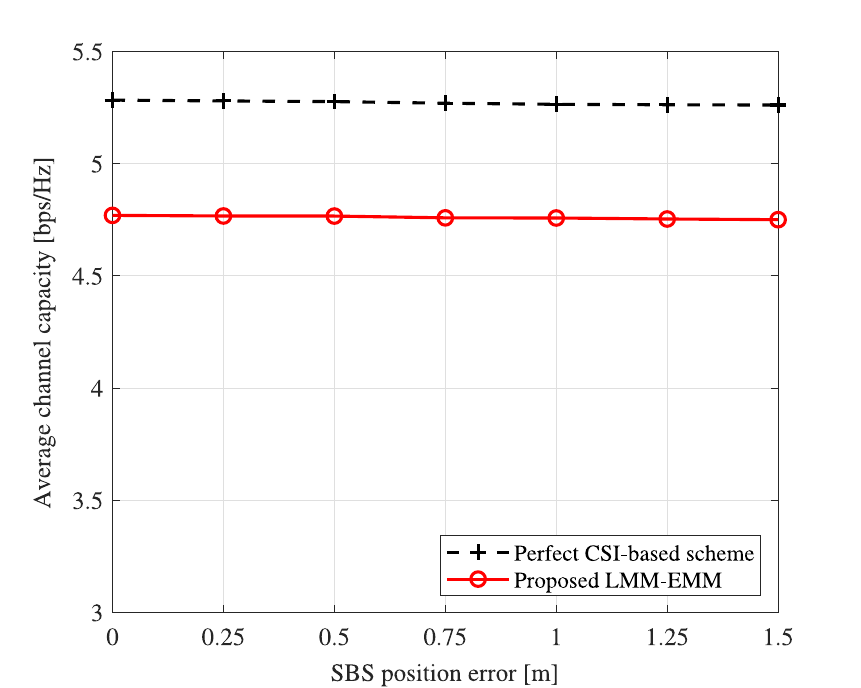}
    \caption{{\color{\blue}Average channel capacity vs. SBS position error.}}
    \label{bs_position}
\end{minipage}
\hfill
\begin{minipage}[t]{0.315\textwidth}
    \centering
    \includegraphics[width=\textwidth]{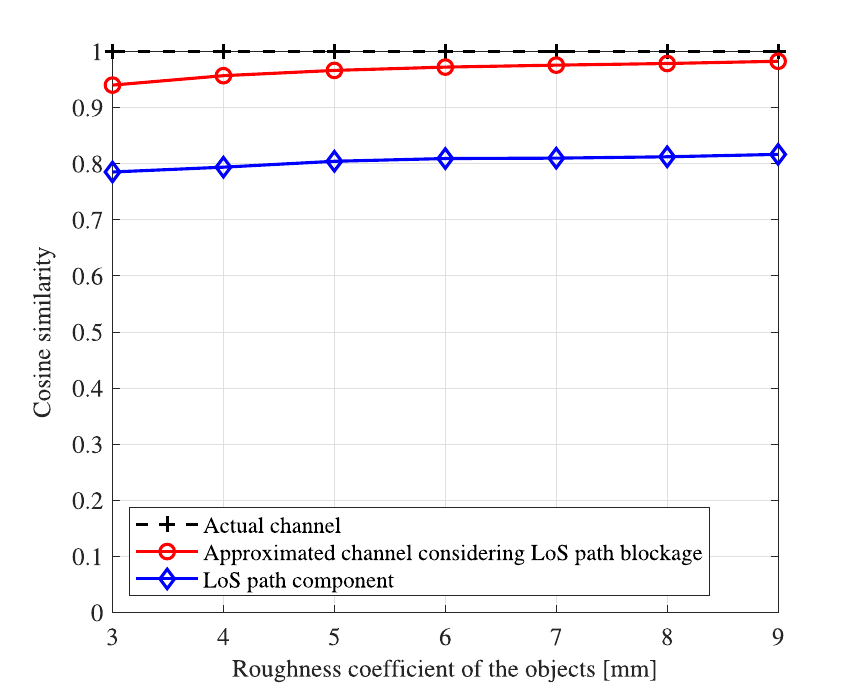}
    \caption{{\color{\blue}Cosine similarity between approximated channel and actual channel.}}
    \label{corr}
\end{minipage}
\vspace{-1em}
\end{figure*}

\subsubsection{SBS Position Error}
Inaccurate \ac{SBS} positions may affect blockage detection performance because the estimated \ac{LoS} path can deviate from the actual one. 
To evaluate the impact of SBS position error, we plot the average channel capacity of the proposed \ac{LMM-EMM} as a function of the SBS position error.
We observe that under an \ac{SBS} position error of $1$\,m, the average channel capacity reaches $4.758$\,bps/Hz, which is only $0.3$\% lower than that under perfect SBS position knowledge (see Fig.~\ref{bs_position}).
Moreover, since SBS positions are typically fixed after deployment, they can be calibrated over time, so persistent location errors are unlikely in practical systems.

\subsubsection{Soft Blockage}
In practical environments, a soft blockage model may arise due to partial blockage and diffraction, which could introduce discrepancies in the channel capacity estimation.
However, in the considered mmWave scenario, the impact of such effects is limited for two main reasons.

First, diffraction and refraction are generally weak at mmWave frequencies due to the short wavelength and high penetration loss~\cite{maccartney2016millimeter}.
As a result, the corresponding path components contribute much less to the received power than the reflected paths.
Their effect on channel capacity variation is thus limited, and neglecting them is a reasonable approximation in the considered mmWave scenario.

Second, the impact of partial blockage on the channel capacity variation is expected to be limited in the considered mmWave scenario.
This is because objects that induce partial blockage (e.g., pedestrians) typically yield a much smaller path gain $\beta_{m,l}[s]$ than specular reflectors (e.g., buildings), as they usually have a significantly larger roughness coefficient $\sigma_{m,l}$ (see Equation~\eqref{pathgain})\footnote{{\color{\blue}For instance, a path partially blocked by a pedestrian with a roughness coefficient of $\sigma_{m,l}=3\,\mathrm{mm}$ yields a $\beta_{m,l}[s]$ that is approximately $1/467$ of that for a typical glass building with $\sigma_{m,l}=1\,\mathrm{mm}$.}}.
Consequently, the channel capacity variation caused by soft blockages is relatively minor compared with the dominant effects of \ac{LoS} and reflected paths.

To validate the adopted \ac{LoS}/\ac{NLoS} approximation combined with the reflection model, we computed the cosine similarity between the approximated and actual channels at various roughness coefficients of the objects. 
We observe that the cosine similarity exceeds $0.93$ even when the roughness coefficient is $3\,\textrm{mm}$, which confirms that the approximated model is sufficiently accurate for channel capacity comparison and handover decision-making (see Fig.~\ref{corr}).

\section{Experimental Results}\label{sec:V}

\subsection{Simulation Setup} \label{sec:V.A}
We consider the \ac{UDN} system where $M=14$ \acp{SBS} equipped with $N_x \times N_y =8 \times 4$ \ac{UPA} antennas serve a \ac{UE} equipped with a single antenna.
The antenna orientations of \ac{SBS} and \ac{UE} are $(0^\circ, -10^\circ, 0^\circ)$ and $(0^\circ, 0^\circ, 0^\circ)$, respectively.
The \ac{UE} moves along the road at a speed of $v=25$\,km/h and it turns left, turns right, or continues straight ahead at each intersection with probabilities of 25\%, 25\%, and 50\%, respectively.
The carrier frequency and bandwidth are $f_{\mathrm{c}}=28$\,GHz and $B=100$\,MHz, respectively.
The observation time window $T_{\mathrm{w}}$ and the prediction length $T_{\mathrm{p}}$ are set to $5$.
We consider the \ac{SNR} of $15$\,dB and the handover interruption time of $\tau_{\mathrm{ho}}=36$\,ms.
\ad{Also, an RGB-D camera having a resolution of $1024 \times 768$ and a \ac{FoV} of $120^\circ$ is mounted on the top side of the \ac{SBS}.}
For the pretrained \ac{LMM}, we adopt the LLaVA-1.5-7B model~\cite{liu2023visual}, which is one of the most prevalent open-source \acp{LMM}.
We conduct experiments on an Intel Xeon Gold 6326 CPU server equipped with a 16-core CPU and an NVIDIA L40S GPU with 48\,GB of memory.


We compare the mobility management performance of the proposed LMM-EMM with four conventional techniques and the ideal scheme with perfect \ac{CSI}:
\begin{enumerate}
    {\color{\blue}
    \item \textit{Perfect \ac{CSI}-based scheme}: an upper bound of average capacity assuming all channel capacities and blockage indicators are perfectly known.
    {\color{\bluetwo}\item \textit{\Ac{DRL}-based scheme}~\cite{sun2025proactive}: A proactive handover scheme that uses \ac{DQN} to optimize the handover parameters (i.e., handover offset and time-to-trigger (TTT)) with handover-type prediction.
    \begin{itemize}
        \item \textbf{State}: \ac{RSRP} and \ac{SINR} from both the serving and target \acp{SBS}, and the \ac{UE}'s distance to the target \ac{SBS}, speed, and moving direction.
        \item \textbf{Action}: handover margin and time-to-trigger.
        \item \textbf{Reward}: negative weighted sum of the ping-pong, too-early, and too-late handover occurrence rates.
    \end{itemize} }
    \item \textit{Vision-aided scheme}~\cite{charan2021vision}: A proactive handover scheme that employs \ac{SBS}-view camera images and a YOLOv3 object detection model to predict the \ac{LoS} path blockages and proactively trigger handover.
    \item \textit{\ac{LSTM}-based scheme}~\cite{shah2022multi}: A proactive handover scheme that uses \ac{LSTM} to predict the future channel capacity in an autoregressive manner and performs handover based on the predicted capacity.
    \item \textit{5G \ac{NR} handover}~\cite{3gpp38331}: A reactive handover scheme where the \ac{SBS} determines the handover decisions based on consecutive \ac{RSRP} reports from the \ac{UE}.
    }
\end{enumerate}

\begin{figure}[t]
        \centering
        \ifthenelse{\equal{\main}{1}}
        {\newcommand{\mywidth}{0.5}}
        {\newcommand{\mywidth}{0.95}}
        \includegraphics[width=\mywidth\linewidth]{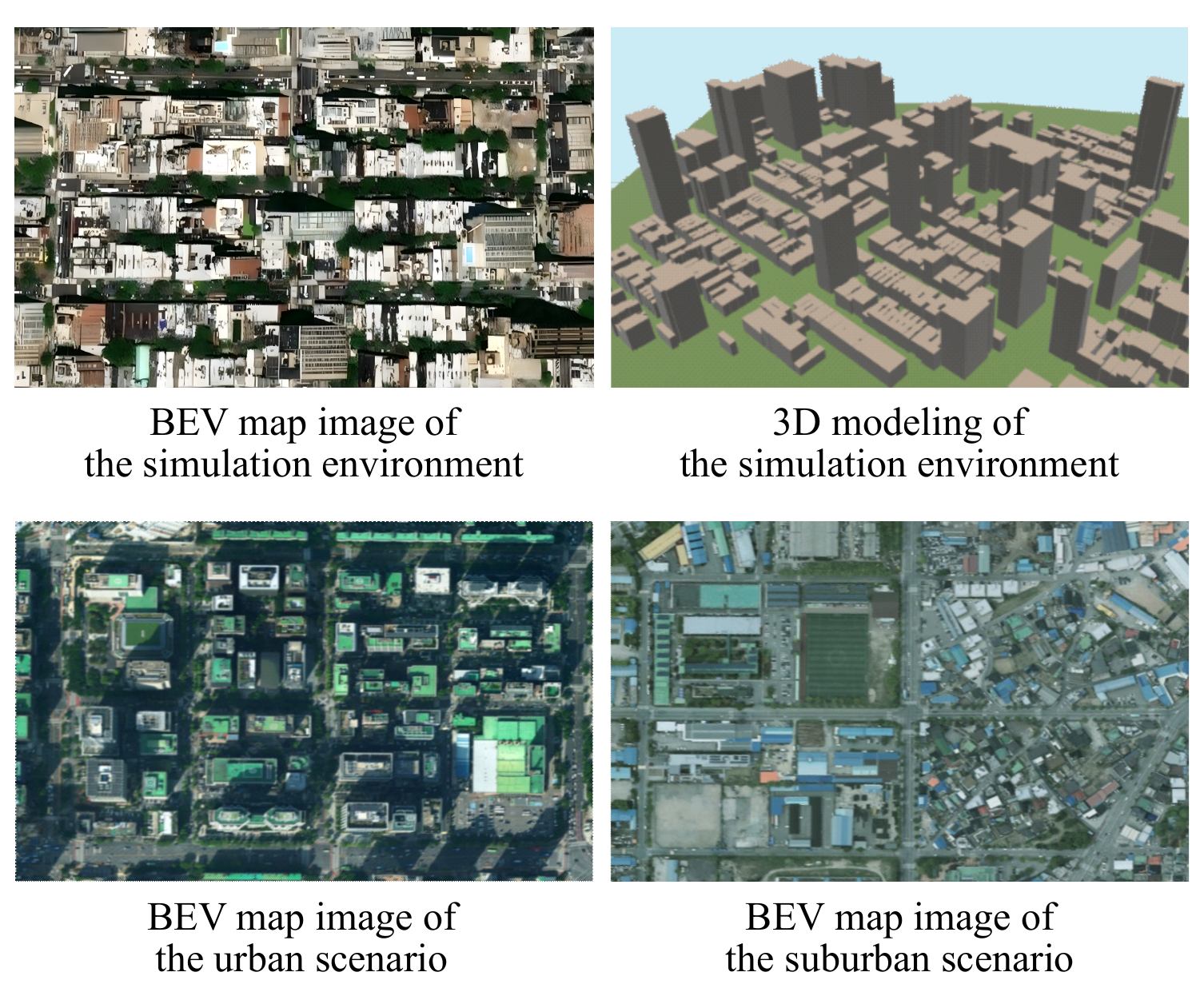}
        \caption{{\color{\blue}Visualizations of the simulation environment on real-world scenarios.}}
        \vspace{-1em}
        \label{fig:8}
\end{figure}

\begin{figure*}[t]
        \centering
        \ifthenelse{\equal{\main}{1}}
        {\newcommand{\mywidth}{0.8}}
        {\newcommand{\mywidth}{0.75}}
        \includegraphics[width=\mywidth\linewidth]{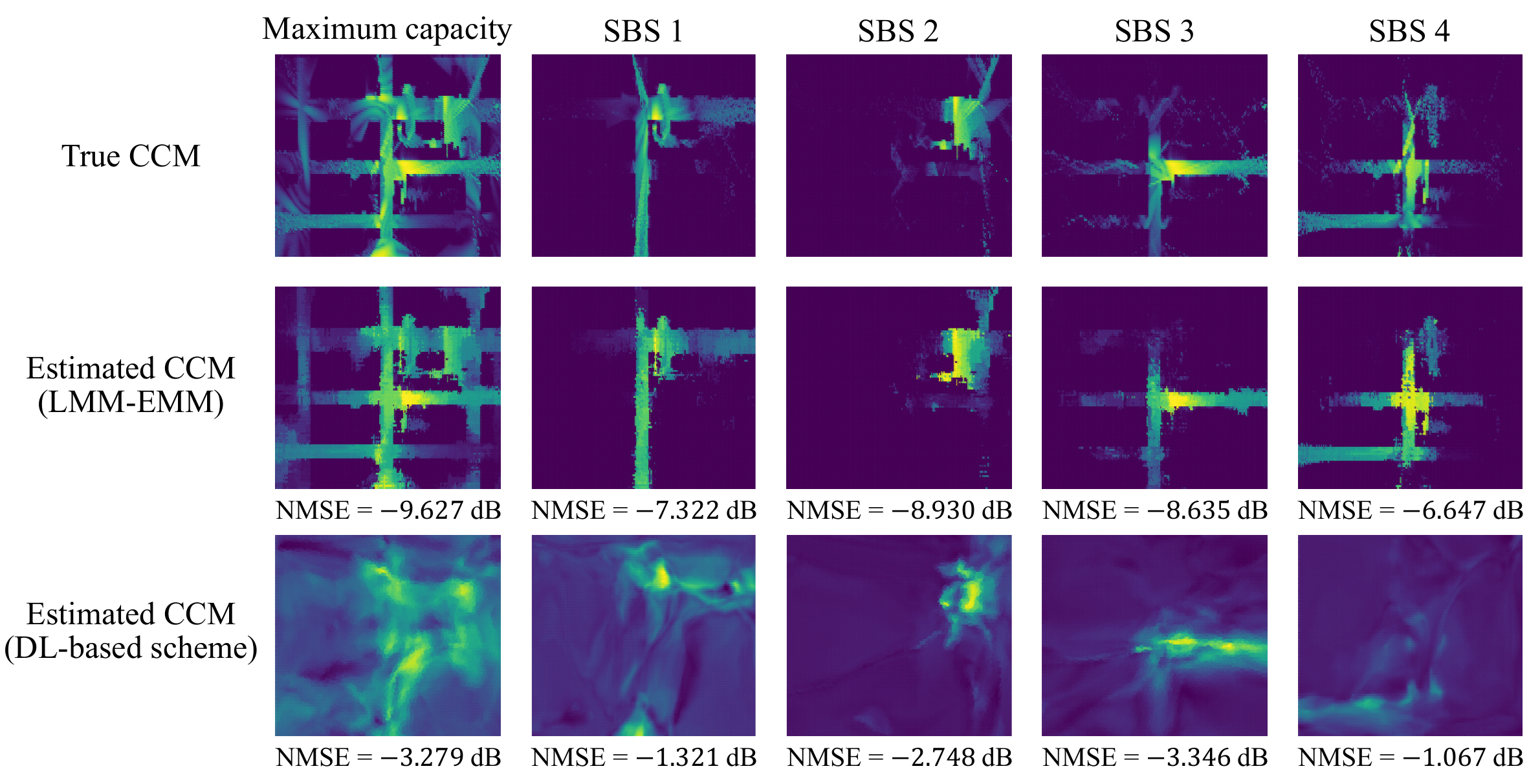}
        \caption{Comparison of \acp{CCM} generated by \ac{LMM-EMM} and the conventional \ac{DL}-based scheme.}
        \vspace{-1em}
        \label{fig:10}
\end{figure*}

\subsection{Multimodal Dataset Generation and LMM Fine-Tuning} \label{sec:V.B}
\subsubsection{Multimodal Dataset Generation}
To construct a realistic dataset for \ac{UDN} scenarios, we develop a comprehensive pipeline that integrates 3D environmental modeling, wireless channel generation, and sensory data acquisition.
\ad{
First, we reconstructed a 3D urban environment resembling a real-world deployment using building layouts extracted from OpenStreetMap, which has been reported to provide sub-meter positional accuracy~\cite{haklay2008openstreetmap}.
Based on this 3D environment, we performed ray tracing and wireless channel generation using NVIDIA Sionna RT~\cite{hoydis2023sionna}, which provides physically consistent channel modeling based on ray tracing and 3GPP TR 38.901~\cite{3gpp38901}.
Since 3GPP TR 38.901 is widely adopted to emulate realistic propagation characteristics (e.g., reflection and blockage), the generated wireless channels closely resemble practical deployments.
Therefore, although the data are simulated, the dataset maintains high physical fidelity and supports reliable performance evaluation for practical deployment scenarios.}
To validate performance in general wireless environments, we generate two additional real-world scenarios: 1) an urban environment characterized by high building density and tall structures, and 2) a suburban environment with moderate building density and low heights (less than $10$\,m) (see Fig.~\ref{fig:8}).

\subsubsection{LMM Fine-Tuning}
\textcolor{\blue}{For \ac{LMM} fine-tuning, we employ \ac{SFT} in which the \ac{LMM} is trained on input–output pairs generated by a real-world wireless simulator (NVIDIA Sionna RT~\cite{hoydis2023sionna}) to minimize the \ac{NLL} loss $\mathcal{J}$ for the response tokens $\boldsymbol{w}^*$:}
\begin{equation}
    \mathcal{J} =
    - \frac{1}{N_{\mathrm{token}}} \sum_{i=1}^{N_{\mathrm{token}}} \log\big([\boldsymbol{\pi}_{i}]_{w_i^*}\big).
\label{loss}
\end{equation}
By adjusting the model parameters through \ac{SFT}, \ac{LMM-EMM} can capture fine-grained mobility behaviors and channel variations, thereby enhancing mobility management performance.
One major issue in \ac{SFT} is the excessive computational burden of updating a massive number of network parameters (e.g., 7 billion parameters in LLaVA-1.5-7B), which results in extended training times and considerable resource demands.
We address this issue by adopting the \ac{LoRA} technique, which adjusts only a small fraction of parameters relevant to the task~\cite{hu2022lora}.
Specifically, we select a subset of parameters $\mathbf{W} \in \mathbb{R}^{X \times Y}$ and update it with the product of two low-rank matrices $\mathbf{A} \in \mathbb{R}^{X \times r}$ and $\mathbf{B} \in \mathbb{R}^{r \times Y}$:
\vspace{-0.3em}
\begin{equation}
    \mathbf{W}' = \mathbf{W} + \mathbf{A}\mathbf{B}
    \vspace{-0.3em}
\end{equation}
where $\mathbf{W}'\in \mathbb{R}^{X \times Y}$ and $r$ denote the updated model parameters and the rank of low-rank matrices, respectively.
\ad{We use a \ac{LoRA} rank of $r = 16$ and a learning rate of $1\times10^{-4}$, which is decayed by a factor of 0.1 every 10 steps.}
\ad{The generated dataset for fine-tuning contains $N_{\mathrm{train}}=16\,000$, $N_{\mathrm{val}}=2\,000$, and $N_{\mathrm{test}}=2\,000$ samples for training, validation, and test, respectively.}
\begin{figure}[t]
        \centering
        \vspace{0.5em}
        \ifthenelse{\equal{\main}{1}}
        {\def\mywidth{0.4\linewidth}}
        {\newcommand{\mywidth}{0.75}}
        \includegraphics[width=0.7\linewidth]{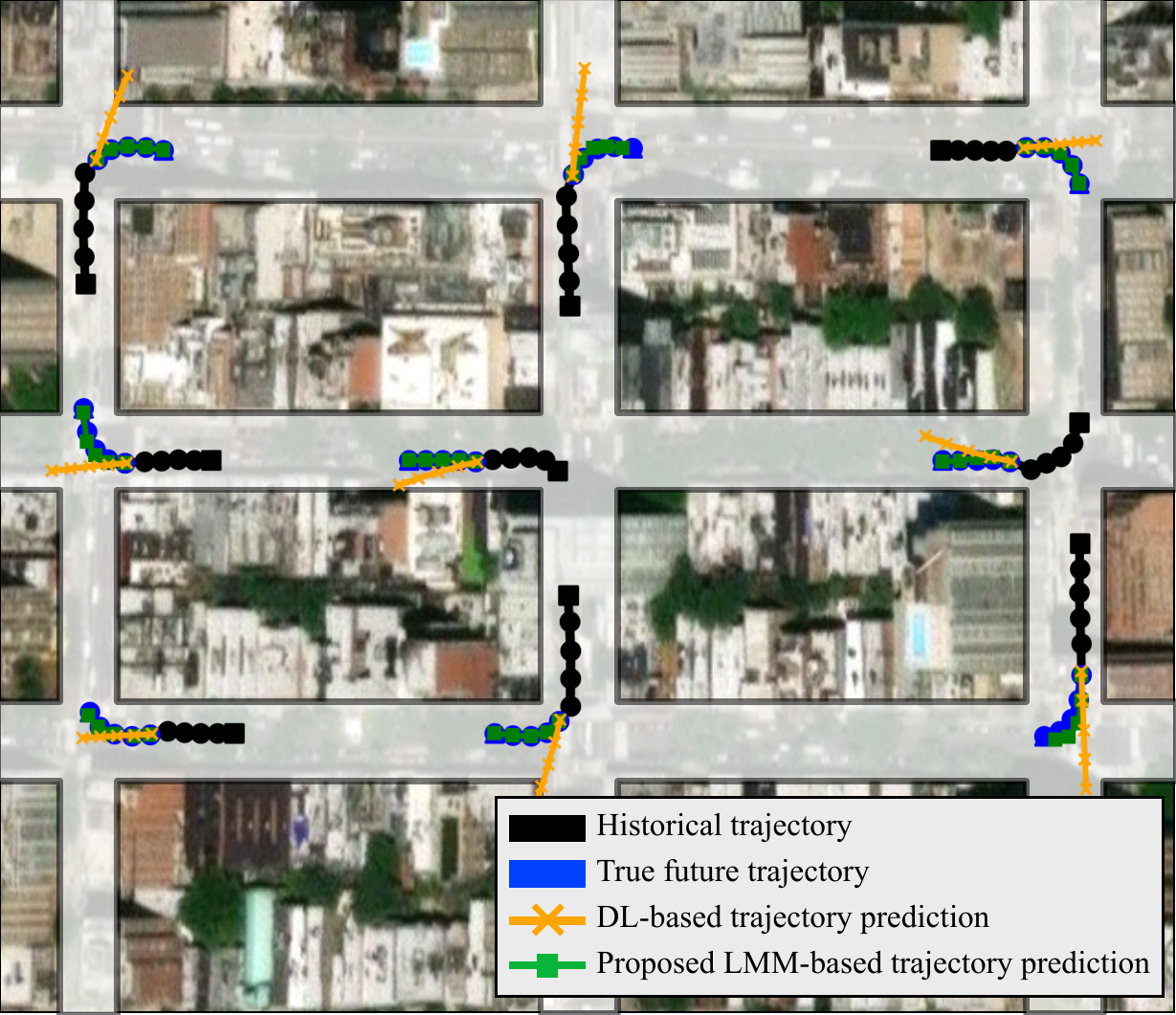}
        \caption{Visual comparison of predicted \ac{UE} trajectories for \ac{LMM-EMM} and the \ac{DL}-based trajectory prediction.}
        \label{fig:11}
        \vspace{-1em} 
\end{figure}

\subsection{Simulation Results} \label{sec:V.C}

In Fig.~\ref{fig:10}, we visualize \ac{CCM} generated by \ac{LMM-EMM} and the conventional \ac{DL}-based scheme~\cite{zeng2021toward}.
To assess the quality of the generated \ac{CCM}, we evaluate the \ac{NMSE} against the ground-truth \ac{CCM} \ad{as shown in Appendix~\hyperref[app:B]{B}.}
The results indicate that the proposed \ac{LMM-EMM} generates a more accurate \ac{CCM} than the \ac{DL}-based scheme, achieving a $6.3$\,dB improvement in \ac{NMSE}.
Additionally, in Fig.~\ref{fig:11}, we compare the predicted \ac{UE} trajectory of \ac{LMM-EMM} and the \ac{DL}-based trajectory prediction.
We observe that the trajectory predicted by \ac{LMM-EMM} closely follows the ground-truth path and accurately captures direction changes and \ac{UE} movement.
Overall, these findings indicate that \ac{LMM-EMM} understands environmental features well, including intersections and obstacle locations, so that it can facilitate reliable handover decisions.

\begin{figure*}[h]
\centering
\begin{minipage}[t]{0.32\textwidth}
    \centering
    \ifthenelse{\equal{\main}{1}}
    {\newcommand{\mywidth}{0.5}}
    {\newcommand{\mywidth}{\graphsize}}
    \includegraphics[width=\mywidth\linewidth]{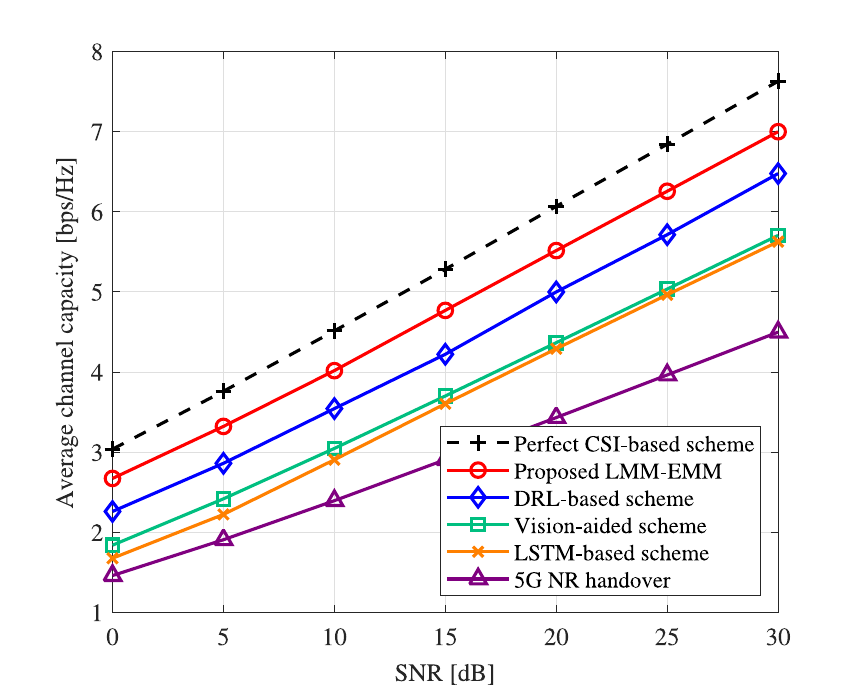}
    \caption{Average channel capacity as a function of \ac{SNR}.}
    \vspace{-1em}
    \label{fig:13}
\end{minipage}
\hfill
\begin{minipage}[t]{0.32\textwidth}
    \centering
    \ifthenelse{\equal{\main}{1}}
    {\newcommand{\mywidth}{0.5}}
    {\newcommand{\mywidth}{\graphsize}}
    \includegraphics[width=\mywidth\linewidth]{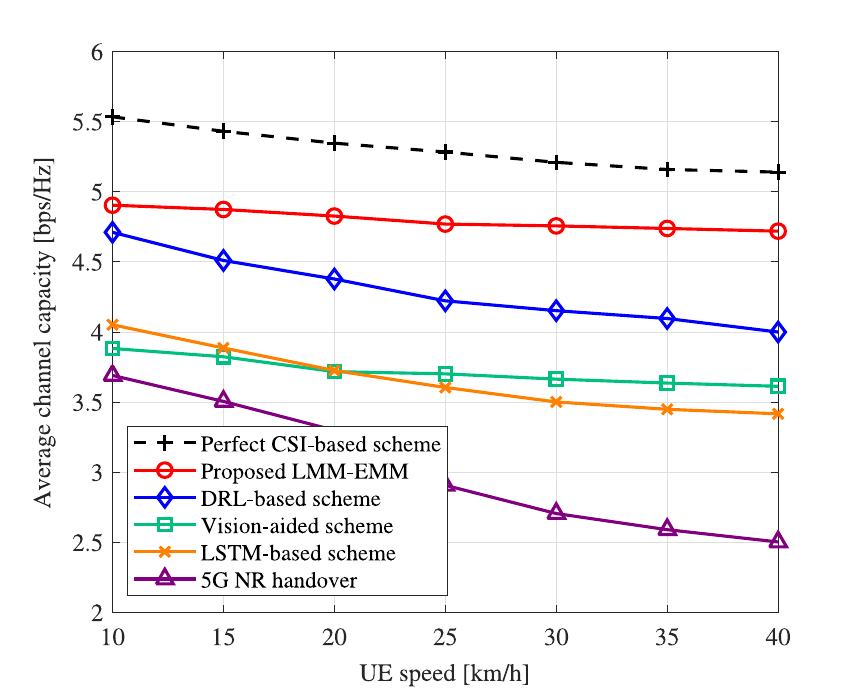}
    \caption{Average channel capacity as a function of the \ac{UE} speed.}
    \vspace{-1em}
    \label{fig:14}
\end{minipage}
\hfill
\begin{minipage}[t]{0.32\textwidth}
    \centering
    \ifthenelse{\equal{\main}{1}}
    {\newcommand{\mywidth}{0.5}}
    {\newcommand{\mywidth}{\graphsize}}
    \includegraphics[width=\mywidth\linewidth]{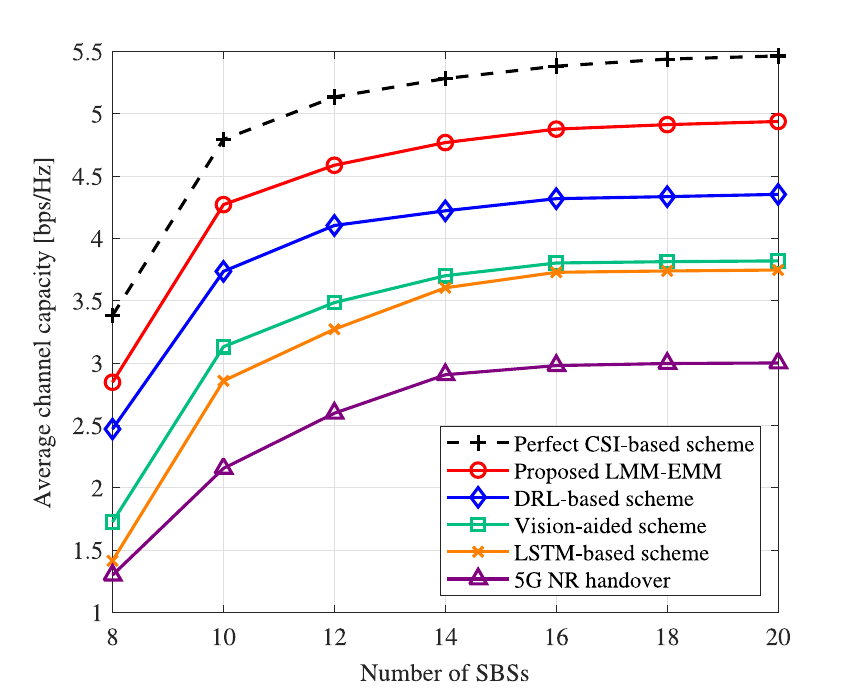}
    \caption{Average channel capacity as a function of the number of \acp{SBS}.}
    \vspace{-1em}
    \label{fig:15}
\end{minipage}
\end{figure*}



In Fig.~\ref{fig:13}, we evaluate the channel capacity as a function of \ac{SNR}.
We see that the proposed \ac{LMM-EMM} achieves a significant improvement in the channel capacity over the conventional mobility management techniques.
For instance, at an \ac{SNR} of 15 dB, \ac{LMM-EMM} achieves more than 45\% gain in channel capacity compared to the 5G \ac{NR} handover scheme.
Even when compared to the \ac{LSTM}-based technique, \ac{LMM-EMM} achieves a 21\% gain in channel capacity.
This is because the proposed \ac{LMM-EMM} can properly comprehend the environmental information to predict the accurate channel capacity from multiple \acp{SBS}.

In Fig.~\ref{fig:14}, we evaluate the handover performance as a function of the UE speed.
Due to the accurate estimation of the future trajectory of the \ac{UE} and corresponding channels from the environment, the proposed \ac{LMM-EMM} outperforms conventional mobility management techniques in terms of channel capacity.
For example, when the speed of the UE is $25$\,km/h, LMM-EMM achieves more than 50\% and 23\% higher channel capacity compared to the 5G \ac{NR} handover and \ac{LSTM}-based technique, respectively.
Since \ac{LMM-EMM} intelligently analyzes road geometry to accurately estimate the \ac{UE} trajectory and corresponding channel capacities, it achieves robust performance across diverse mobility scenarios.




To evaluate the performance of \ac{LMM-EMM} in various deployment scenarios, we plot the channel capacity of the UE as a function of the number of \acp{SBS} in Fig.~\ref{fig:15}.
Since each \ac{UDN} system contains a different number of \acp{SBS}, validation across diverse numbers of \acp{SBS} is crucial to demonstrate generalization performance.
We see that the channel capacity of the proposed \ac{LMM-EMM} is much higher than that of conventional techniques.
For example, when the number of \acp{SBS} is $12$, \ac{LMM-EMM} achieves more than 40\% higher capacity compared to the 5G \ac{NR} handover.
By exploiting \ac{CCM}, \ac{LMM-EMM} can accurately estimate the channel capacity of each \ac{SBS} and associate the appropriate \ac{SBS} with the \ac{UE}, regardless of the number of \acp{SBS}.

In Fig.~\ref{fig:16}, we evaluate the handover performance with various handover interruption times.
The results demonstrate that the proposed \ac{LMM-EMM} significantly outperforms conventional mobility management techniques.
For instance, when $\tau_{\mathrm{ho}}=54$\,ms, \ac{LMM-EMM} achieves more than 38\% capacity gain over the 5G \ac{NR} handover.
Even when compared to the vision-based technique, \ac{LMM-EMM} achieves a 17\% increase in channel capacity.
Notably, \ac{LMM-EMM} exhibits minimal degradation in average capacity as the handover interruption time increases.
This indicates that \ac{LMM-EMM} can avoid redundant handovers through accurate recognition of transient channel variations caused by dynamic obstacles.

To assess the generalization performance of \ac{LMM-EMM}, we evaluate the channel capacity across diverse scenarios, including urban and suburban areas, in Fig.~\ref{fig:17}.
We observe that \ac{LMM-EMM} exhibits superior performance against conventional techniques across all wireless scenarios.
For example, \ac{LMM-EMM} achieves 30\% and 26\% higher channel capacity than the \ac{DRL}-based technique in urban and suburban settings, respectively.
This generalization performance results from the multimodal reasoning capability of \ac{LMM}, which accurately interprets the spatial distribution of reflectors and the \ac{UE} mobility patterns in diverse environments.



\begin{figure*}[h]
\centering
\begin{minipage}[t]{0.32\textwidth}
    \centering
    \ifthenelse{\equal{\main}{1}}
    {\newcommand{\mywidth}{0.5}}
    {\newcommand{\mywidth}{\graphsize}}
    \includegraphics[width=\mywidth\linewidth]{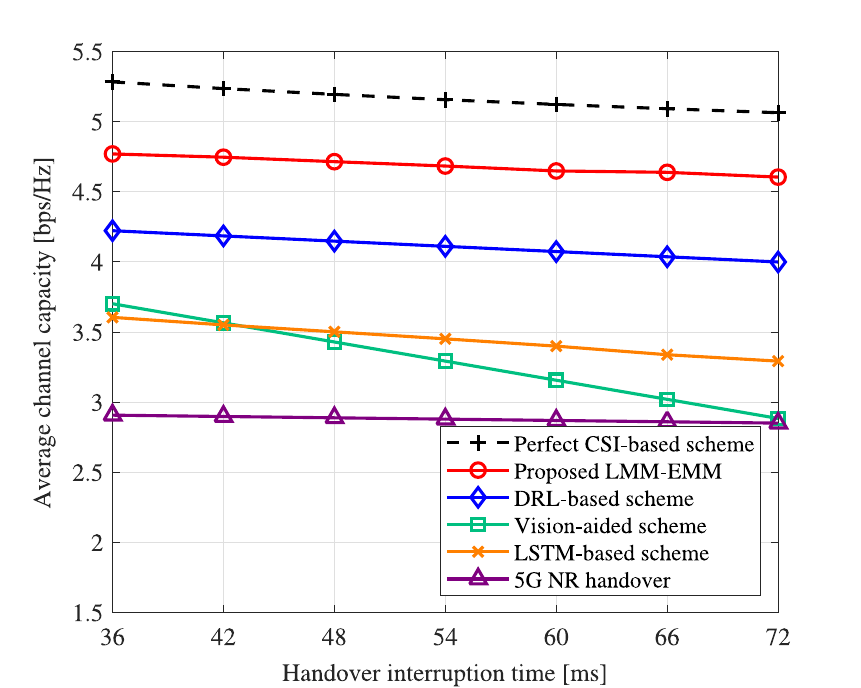}
    \caption{Average channel capacity as a function of handover interruption time.}
    \label{fig:16}
\end{minipage}
\hfill
\begin{minipage}[t]{0.32\textwidth}
    \centering
    \ifthenelse{\equal{\main}{1}}
    {\newcommand{\mywidth}{0.5}}
    {\newcommand{\mywidth}{\graphsize}}
    \includegraphics[width=\mywidth\linewidth]{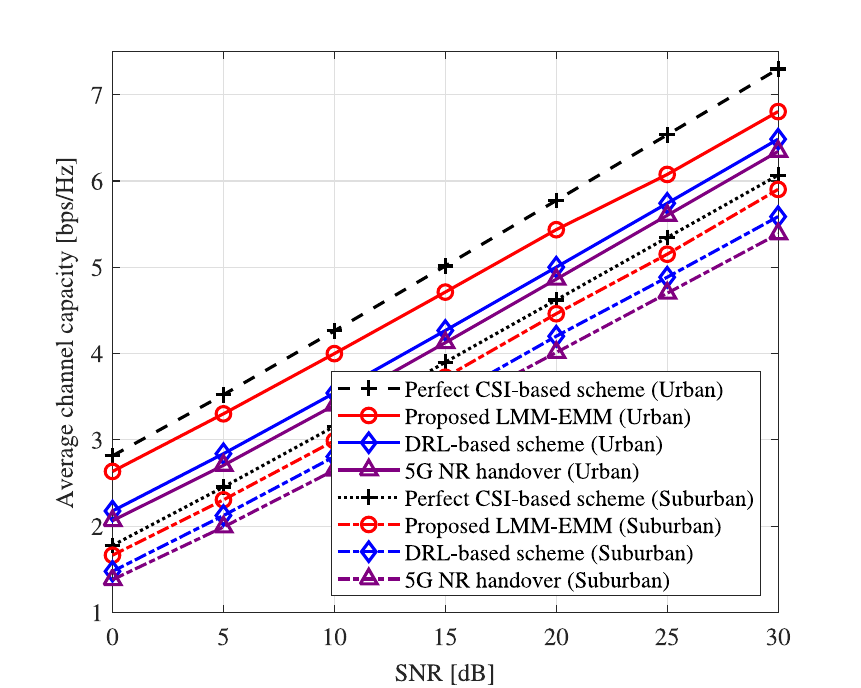}
    \caption{Average channel capacity as a function of \ac{SNR} in various wireless scenarios.}
    \label{fig:17}
\end{minipage}
\hfill
\begin{minipage}[t]{0.32\textwidth}
    \centering
    \ifthenelse{\equal{\main}{1}}
    {\newcommand{\mywidth}{0.5}}
    {\newcommand{\mywidth}{\graphsize}}
    \includegraphics[width=\mywidth\linewidth]{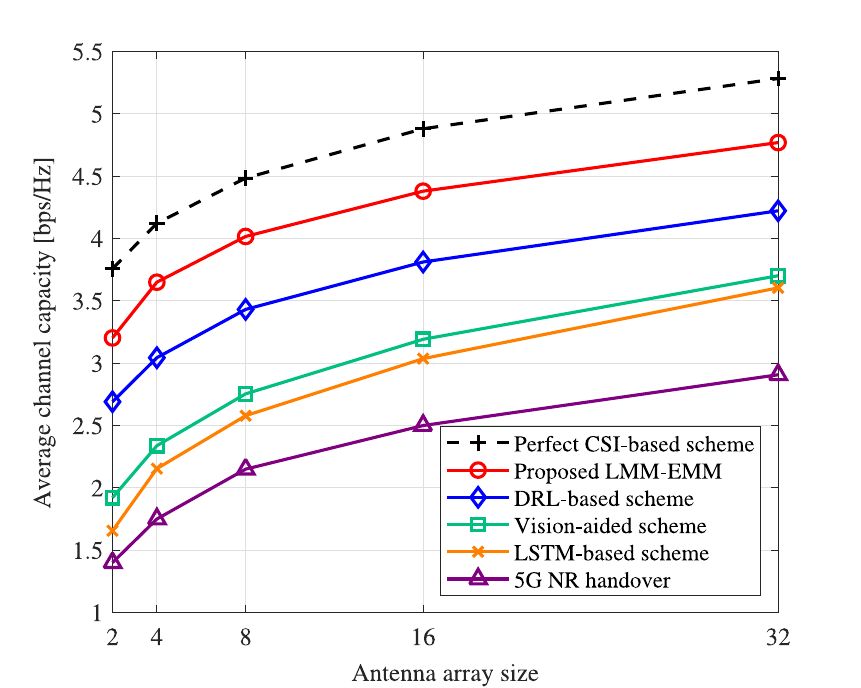}
    \caption{{\color{\blue}Average channel capacity as a function of antenna array size.}}
    \label{antenna}
\end{minipage}
\end{figure*}



{\color{\blue}
To verify that \ac{LMM-EMM} remains effective even with smaller antenna arrays, in Fig.~\ref{antenna}, we evaluate the channel capacity under various antenna array sizes.
We observe that although the instantaneous channel experiences fluctuations due to small-scale fading, these fluctuations rarely change the handover decision.
For example, even when the antenna array size is $8$, LMM-EMM achieves $4.017\,$bps/Hz, which corresponds to $90\%$ of the channel capacity achieved by the optimal handover with perfect CSI.
These results indicate that the proposed LMM-EMM retains most of the ideal handover gain even when the number of antennas is small.

In Fig.~\ref{fewshot}, we evaluate the average channel capacity performance of \ac{LMM-EMM} across different architectural layouts, including highway and indoor scenarios.
To this end, we perform few-shot fine-tuning using only $600$ samples collected from those environments.
We observe that at an SNR of $15\,$dB, LMM-EMM still achieves $10$\% and $7$\% channel capacity gain over the conventional \ac{DRL}-based approach in highway and indoor scenarios, respectively (see Fig.~\ref{fewshot}).
This strong adaptability of \ac{LMM-EMM} stems from the ability of \ac{LMM} to capture the scenario-invariant propagation mechanisms (e.g., reflection and blockage) that are shared across different environments.
While architectural layouts vary across environments, these variations primarily affect site-specific geometry, not the fundamental propagation physics.
This means that \ac{LMM} can quickly understand the propagation characteristics in new environments and rapidly adapt to unseen scenarios.
}

\begin{table*}[t]
    \ifthenelse{\equal{\main}{1}}
    {\renewcommand{\arraystretch}{1.2}}
    {\renewcommand{\arraystretch}{1.4}}
    \centering
    \caption{{\color{\bluetwo}Impact of input UE trajectory errors on prediction accuracy and communication performance.}}
    \begin{tabular}{c | ccccc}
        \hline
        \textbf{Noise variance} & \textbf{$0\,$m} & \textbf{$0.5\,$m} & \textbf{$1\,$m} & \textbf{$1.5\,$m} & \textbf{$2\,$m} \\
        \hline
        \rowcolor{blue!20!white}
        \textbf{UE trajectory prediction RMSE} 
        & $0.252\,$m 
        & $0.490\,$m   
        & $0.675\,$m 
        & $0.844\,$m  
        & $0.964\,$m  \\
        
        \rowcolor{blue!0!white}
        \textbf{Static channel capacity estimation NMSE} 
        & $-9.627\,$dB 
        & $-9.327\,$dB 
        & $-8.549\,$dB 
        & $-7.542\,$dB 
        & $-6.379\,$dB  \\

        \rowcolor{blue!20!white}
        \textbf{Blockage prediction accuracy} 
        & $95.5\%$ 
        & $95.4\%$ 
        & $94.9\%$ 
        & $93.7\%$  
        & $92.2\%$ \\

        \rowcolor{blue!0!white}
        \textbf{Average channel capacity} 
        & $4.770\,$bps/Hz 
        & $4.729\,$bps/Hz 
        & $4.657\,$bps/Hz 
        & $4.556\,$bps/Hz 
        & $4.371\,$bps/Hz \\
        
        \rowcolor{blue!20!white}
        \textbf{10\% worst case channel capacity} 
        & $2.490\,$bps/Hz 
        & $2.479\,$bps/Hz 
        & $2.440\,$bps/Hz 
        & $1.987\,$bps/Hz 
        & $1.798\,$bps/Hz \\

        \rowcolor{blue!0!white}
        \textbf{20\% worst case channel capacity} 
        & $2.744\,$bps/Hz 
        & $2.736\,$bps/Hz 
        & $2.670\,$bps/Hz 
        & $2.363\,$bps/Hz 
        & $2.181\,$bps/Hz \\

        \rowcolor{blue!20!white}
        \textbf{Handover failure rate} 
        & $4.89\,$\% 
        & $5.98\,$\% 
        & $6.52\,$\% 
        & $8.42\,$\% 
        & $9.24\,$\% \\

        \rowcolor{blue!0!white}
        \textbf{Service interruption probability} 
        & $6.61\,$\% 
        & $6.81\,$\% 
        & $6.90\,$\% 
        & $7.25\,$\% 
        & $7.82\,$\% \\
        \hline
    \end{tabular}
    \label{ablation}
\end{table*}

\begin{table*}[t]
    \ifthenelse{\equal{\main}{1}}
    {\renewcommand{\arraystretch}{1.2}}
    {\renewcommand{\arraystretch}{1.4}}
    \centering
    \caption{{\color{\bluetwo}Ablation study on the impact of UE trajectory prediction, CCM estimation, and blockage prediction error}}
    \begin{tabular}{cc | ccc}
        \hline
        \multirow{2}{*}{\textbf{Error source}}   & \multirow{2}{*}{\textbf{Amount of error}} & \textbf{UE trajectory} & \textbf{Channel capacity} & \textbf{Average} \\
        & & \textbf{prediction RMSE} & \textbf{estimation NMSE} &  \textbf{channel capacity} \\
        \hline
        \rowcolor{blue!20!white} 
        No noise addition &
        - &
        $0.252\,$m & 
        $-9.627\,$dB & 
        $4.770\,$bps/Hz \\
        
		\rowcolor{blue!0!white} 
        Input UE trajectory & 
        Noise variance = $1\,$m & 
        $0.675\,$m & 
        $-9.061\,$dB & 
        $4.659\,$bps/Hz \\
        
        \rowcolor{blue!20!white} 
        UE trajectory prediction & 
        Noise variance = $1\,$m & 
        $1.083\,$m & 
        $-5.853\,$dB & 
        $4.632\,$bps/Hz\\
        
        \rowcolor{blue!0!white}
        CCM estimation & 
        Noise variance = $3\,$bps/Hz & 
        - & 
        $-4.680\,$dB & 
        $4.554\,$bps/Hz\\
        
        \rowcolor{blue!20!white} 
        Blockage prediction & 
        Error probability +10\%p & 
        - & 
        - & 
        $4.747\,$bps/Hz\\
        \hline 
        \end{tabular}  
    \label{propagation}
\end{table*}

{\color{\bluetwo}
Fig.~\ref{cnn} shows an ablation study where each module is replaced with a CNN+LSTM model.
Replacing the trajectory prediction, channel capacity estimation, and blockage prediction modules with CNN+LSTM leads to data-rate degradations of $6.4$\%, $24.4$\%, and $6.1$\%, respectively, since a simple discriminative model lacks the multimodal reasoning provided by \ac{LMM}.
The degradation is particularly noticeable for the channel capacity estimation module, because the channel capacity is more sensitive to variations and piecewise discontinuities in the propagation environment than the UE and blockage trajectories.
Also, in Fig.~\ref{tinyllava}, we conduct an experiment where the backbone LMM is replaced with lightweight models: 1) MobileVLM-1.7B~\cite{chu2023mobilevlm} and 2) TinyLLaVA-1.5B~\cite{zhou2024tinyllava}.
We observe that MobileVLM-1.7B and TinyLLaVA-1.5B achieve $4.263$\,bps/Hz and $4.305$\,bps/Hz, respectively.
Despite reducing the parameter size by more than 75\% compared with LLaVA-1.5-7B, the performance degradation remains marginal, which demonstrates the feasibility of lightweight deployment.
}

{\color{\blue}
\adtwo{To investigate how the accumulation of errors affects overall system performance, we conduct an error propagation experiment across the entire framework in Table~\ref {ablation}.
Specifically, we first inject artificial noise into the input \ac{UE} trajectory and then evaluate its impact on various metrics, including the \ac{UE} trajectory prediction \ac{RMSE}, static channel capacity estimation \ac{NMSE}, average channel capacity, and the worst $10\%$ and $20\%$ case channel capacity.
The definitions of \ac{RMSE} and \ac{NMSE} are provided in Appendix~\hyperref[app:B]{B}.}
When noise with variance $1\,$m is added to the input UE trajectory, the UE trajectory prediction RMSE is $0.675\,$m and the average channel capacity is $4.657\,$bps/Hz.
Compared with the noise-free case, the channel capacity decreases by only $2.4\%$, indicating that the performance degradation is limited.
}

\adtwo{Table~\ref{propagation} presents the robustness of LMM-EMM to the error propagation in real-world scenarios.
The error level for each module was chosen to reflect practically plausible conditions, motivated by prior studies~\cite{italiano2024tutorial,nishio2019proactive,alrabeiah2020deep}.
As shown in Table II, the performance is most sensitive to errors in the CCM estimation module. 
When the noise with a variance of $3$\,bps/Hz is added to the CCM estimation, the cumulative capacity per unit bandwidth decreases to $4.554$\,bps/Hz, which is the lowest among all error sources. 
This is because the noise in the CCM makes it difficult for \ac{LMM} to learn the underlying physical relationship between the propagation environment (e.g., reflection) and estimate the resulting channel capacity.}

\adtwo{In real-world scenarios, small-scale elements (e.g., trees, small objects) can cause temporary blockages and scattering effects, leading to some minor errors in blockage detection and channel capacity estimation.
To account for such real-world variability, in Table~\ref{noise}, we evaluate the average channel capacity under additional blockage detection errors and CCM estimation noise.
We observe that even with an additional blockage error of 20\% and a CCM noise standard deviation of $2\,$bps/Hz, LMM-EMM achieves an average channel capacity of $4.659\,$bps/Hz, corresponding to only a $2.4\%$ performance degradation.
This robustness is because handover decisions are often dominated by user location and major blockage events caused by large reflectors or obstacles, rather than by minor dispersion effects from small-scale objects.}

\begin{figure*}[h]
\centering
\begin{minipage}[t]{0.32\textwidth}
    \centering
    \ifthenelse{\equal{\main}{1}}
    {\newcommand{\mywidth}{0.5}}
    {\newcommand{\mywidth}{\graphsize}}
    \includegraphics[width=\mywidth\linewidth]{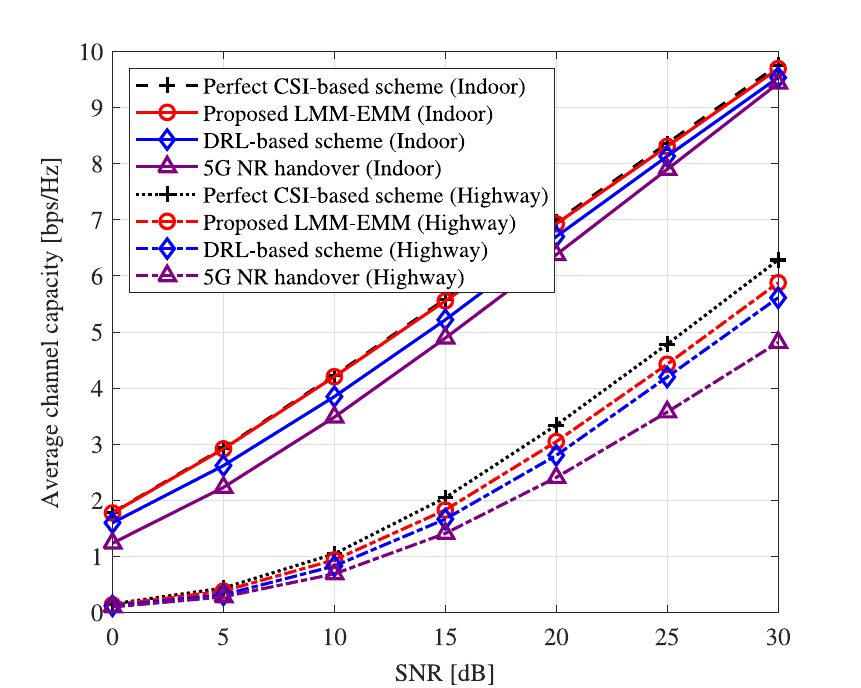}
    \caption{{\color{\blue}Average channel capacity as a function of \ac{SNR} in different architectural layouts after few-shot adaptation.}}
    \label{fewshot}
\end{minipage}
\hfill
\begin{minipage}[t]{0.32\textwidth}
    \centering
    \ifthenelse{\equal{\main}{1}}
    {\newcommand{\mywidth}{0.5}}
    {\newcommand{\mywidth}{\graphsize}}
    \includegraphics[width=\mywidth\linewidth]{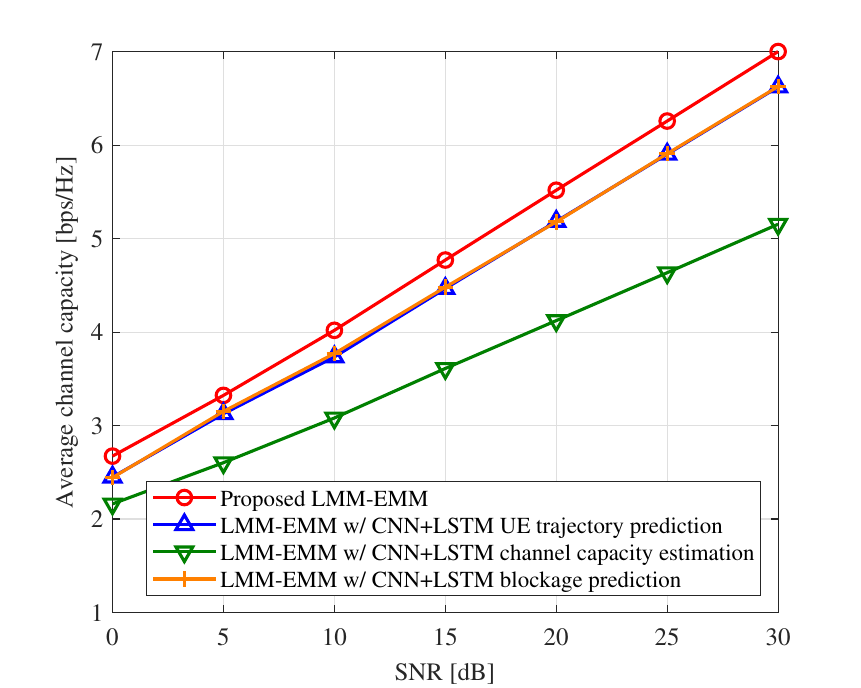}
    \caption{{\color{\bluetwo}Average channel capacity vs. SNR, where each module is replaced by a CNN+LSTM.}}
    \label{cnn}
\end{minipage}
\hfill
\begin{minipage}[t]{0.32\textwidth}
    \centering
    \ifthenelse{\equal{\main}{1}}
    {\newcommand{\mywidth}{0.5}}
    {\newcommand{\mywidth}{\graphsize}}
    \includegraphics[width=\mywidth\linewidth]{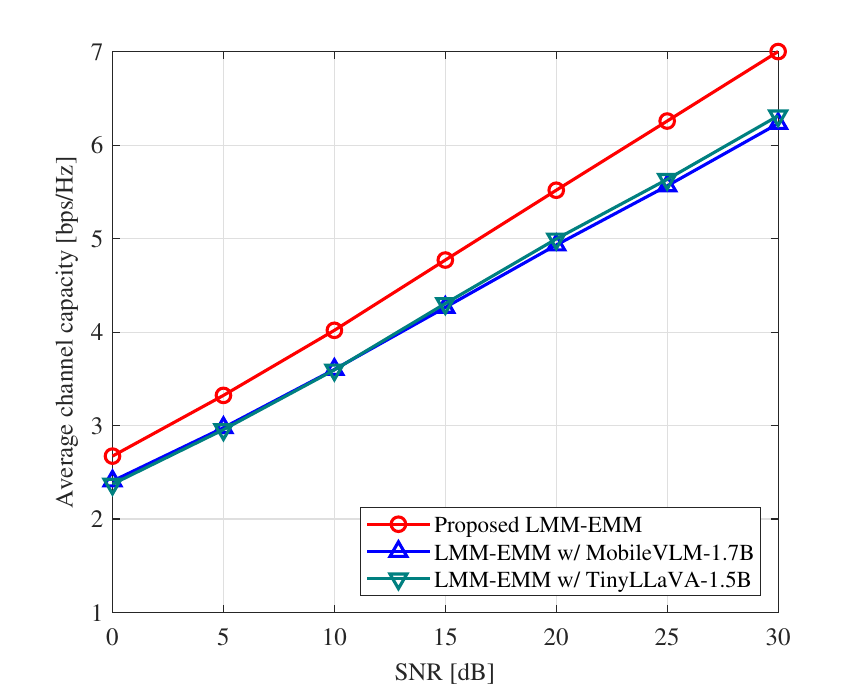}
    \caption{{\color{\bluetwo}Average channel capacity vs. SNR with the backbone replaced by a lightweight LMM.}}
    \label{tinyllava}
\end{minipage}
\vspace{-1em}
\end{figure*}

In Table~\ref{tab:1}, we present the memory and computational complexities of \ac{LMM-EMM} compared to conventional mobility management techniques.
Although \ac{LMM-EMM} exhibits the highest memory usage and longest inference time, its inference time remains shorter than the maximum handover latency (i.e., $360\,$ms) specified in 5G \ac{NR}.
Furthermore, the results can be pre-computed in advance of actual handover events by predicting future channel capacities, which further relaxes the practical constraints on inference time.
This indicates that \ac{LMM-EMM} has a feasible time complexity and is therefore compatible with the 5G \ac{NR} standard.
Moreover, recent progress in lightweight \ac{LMM} and GPU hardware is expected to further alleviate memory and time complexities, so these issues are unlikely to pose practical concerns~\cite{chen2024efficient}.

\section{Conclusion}\label{sec:VI}
In this paper, we proposed an environment-aware mobility management scheme for mmWave \ac{UDN} systems.
The key idea of the proposed \ac{LMM-EMM} scheme is to leverage the \ac{CCM}, an intrinsic mapping from \ac{UE} and \ac{SBS} positions to channel capacity.
By harnessing the multimodal reasoning capability of \acp{LMM}, \ac{LMM-EMM} captures the mobility pattern of the \ac{UE} as well as reflection geometry and transient blockages caused by obstacles. 
Using the trained \ac{CCM} with the predicted \ac{UE} trajectory and blockage indicators, \ac{LMM-EMM} estimates future channel capacities and employs \ac{DP} to proactively determine handover decisions that maximize cumulative channel capacity.
From numerical evaluations on various environments, we demonstrated that \ac{LMM-EMM} achieves substantial channel capacity gains over the conventional \ac{DL}-based methods.
In this paper, we restricted our attention to mobility management but we believe that there are many research extensions of \ac{LMM-EMM} such as user scheduling and random access.

\begin{table}[t]
    \ifthenelse{\equal{\main}{1}}
    {\renewcommand{\arraystretch}{1.2}}
    {\renewcommand{\arraystretch}{1.4}}
    \centering
    \caption{{\color{\bluetwo} Average channel capacity (bps/Hz) at SNR $=15$\,dB, with blockage detection error rate $p_{\mathrm{block}}$ and CCM noise standard deviation $\sigma_n$.}}
    \begin{tabular}{c | ccccc}
        \hline
        \textbf{$\sigma_n  \backslash  p_{\mathrm{block}}$} & \textbf{10\%} &  \textbf{15\%} & \textbf{20\%} & \textbf{25\%} & \textbf{30\%} \\
        \hline
        \rowcolor{blue!20!white}
        \textbf{0.5\,bps/Hz} 
        & $4.754$
        & $4.754$
        & $4.743$ 
        & $4.738$ 
        & $4.727$ \\
        
        \rowcolor{blue!0!white}
        \textbf{1\,bps/Hz} 
        & $4.736$
        & $4.735$
        & $4.726$ 
        & $4.721$ 
        & $4.706$ \\

        \rowcolor{blue!20!white}
        \textbf{1.5\,bps/Hz} 
        & $4.706$
        & $4.704$
        & $4.697$ 
        & $4.695$ 
        & $4.678$ \\

        \rowcolor{blue!0!white}
        \textbf{2\,bps/Hz} 
        & $4.665$
        & $4.668$
        & $4.659$ 
        & $4.662$ 
        & $4.641$ \\
        
        \rowcolor{blue!20!white}
        \textbf{2.5\,bps/Hz} 
        & $4.642$
        & $4.647$
        & $4.647$ 
        & $4.653$ 
        & $4.616$ \\
        \hline
    \end{tabular}
    \label{noise}
\end{table}

\vspace{3em}
\section*{Appendix A\\Proof of Lemma~\ref{lemma:1}} \label{app:A}
In \ac{MISO} systems, the channel capacity is expressed as
\begin{IEEEeqnarray}{rCl}
         R^{(t)}(m)
         &=& B\sum_{s=1}^{S}\log_{2} \bigg( 
         1+\frac{P_{\mathrm{t}}}
         {N \sigma_{\mathrm{n}}^{2}} 
         \sum_{i=1}^{N}\vert h_{m}^{(t)}[s, i]\vert^2
         \bigg) \IEEEeqnarraynumspace \label{capacity_hk}
\end{IEEEeqnarray}
where $h_{m}^{(t)}[s, i]$ is the $i$th element of $\mathbf{h}_{m}^{(t)}[s]$.
When $N$ is sufficiently large, by applying the law of large numbers, we obtain
\begin{IEEEeqnarray}{rCl}
    \frac{1}{N} 
    \sum_{i=1}^{N}
    \vert h_{m}^{(t)}[s, i]\vert^2
    &\approx& \mathbb{E}\big[\vert h_{m}^{(t)}[s, i]\vert^2\big] \\ 
    & =& \sum_{l=0}^{L-1}
    \beta_{m,l}^{(t)} \mathbb{E}\big[\vert\alpha_{m,l}^{(t)}\vert^2\big]
      \notag\\
    &&+\sum_{p \neq l}
    \sqrt{\beta_{m,p}^{(t)}\beta_{m,l}^{(t)}}
    \mathbb{E} \big[
        \alpha_{m,p}^{(t)} \bar{\alpha}_{m,l}^{(t)} 
    \big] e^{j\vartheta}\IEEEeqnarraynumspace \\
    &=& \beta_{m,0}^{(t)} + \sum_{l=1}^{L-1}
    \beta_{m,l}^{(t)}. \label{beta}
\end{IEEEeqnarray}
This is from $\mathbb{E}\big[\vert\alpha_{m,l}^{(t)}\vert^2\big]$$\,=\,$$1$ and $\mathbb{E} \big[ \alpha_{m,p}^{(t)} \bar{\alpha}_{m,l}^{(t)} \big]$$\,=\,$$0$ for $p \neq l$, where $\bar{\alpha}_{m,l}^{(t)}$ is the conjugate of ${\alpha}_{m,l}^{(t)}$ and $\vartheta$ is some real value between $0$ and $2\pi$.
By plugging~\eqref{beta} into~\eqref{capacity_hk}, we finally obtain
\begin{equation}
     R^{(t)}(m) = B\sum_{s=1}^{S} \log_{2} \bigg( 
     1+\frac{P_{\mathrm{t}}}
     {\sigma_{\mathrm{n}}^{2}} 
     \bigg( \beta_{m,0}^{(t)} + \sum_{l=1}^{L-1}\beta_{m,l}^{(t)} 
     \bigg) \bigg).
\end{equation}

\begin{table}[t]
    \ifthenelse{\equal{\main}{1}}
    {\renewcommand{\arraystretch}{1.2}}
    {\renewcommand{\arraystretch}{1.4}}
    \centering
    \caption{Memory usage and inference time}
    \begin{tabular}{cccc}
        \hline
        \textbf{Methods} &\textbf{Memory usage} & \textbf{Inference time}\\
        \hline
		\rowcolor{blue!20!white} 
        Proposed LMM-EMM& $14.16\,$GB & $195.72\,$ms \\
        \rowcolor{blue!0!white}
        \ac{DRL}-based scheme& $1.24\,$GB & $45.68\,$ms \\
        \rowcolor{blue!20!white} 
        Vision-aided scheme&$0.82\,$GB& $14.91\,$ms \\
        \rowcolor{blue!0!white}
        \ac{LSTM}-based scheme& $0.27\,$GB & $6.69\,$ms \\
        \hline 
        \end{tabular}  
    \label{tab:1}
    \vspace{-1em}
\end{table}

{\color{\blue}
\section*{Appendix B\\Evaluation Metrics} \label{app:B}
This appendix defines the evaluation metrics used in this paper, namely the \ac{RMSE} for \ac{UE} trajectory prediction and the \ac{NMSE} for channel capacity estimation. 
\begin{enumerate}
    \item \textit{\ac{UE} trajectory prediction \ac{RMSE}}:
    The \ac{RMSE} for \ac{UE} trajectory prediction is defined as
    \begin{equation}
    \mathrm{RMSE} = \sqrt{\mathbb{E} \Big[\Big(\hat{\mathbf{p}}_{\mathrm{ue}}^{(T:T+T_p)} - \mathbf{p}_{\mathrm{ue}}^{(T:T+T_p)}\Big)^2  \Big]}
    \end{equation}
    where $\hat{\mathbf{p}}_{\mathrm{ue}}^{(T:T+T_p)}$ is the estimated future \ac{UE} trajectory over the prediction horizon from $T$ to $T+T_p$.
    
    \item \textit{Channel capacity estimation \ac{NMSE}}:
    The \ac{NMSE} for channel capacity estimation is defined as
    \begin{equation}
    \mathrm{NMSE} = \mathbb{E}\bigg[ \frac{\big(\hat{R}^{(t)}(m) - {R}^{(t)}(m)\big)^2}{R^{(t)}(m)^2} \bigg]
    \end{equation}
    where $\hat{R}^{(t)}(m)$ is the estimated channel capacity at the $t$th time slot for the $m$th \ac{SBS}.
\end{enumerate}
}

\bibliography{LatexInclusions/References}
\begin{IEEEbiography}[{\includegraphics[width=1in,height=1.35in,clip,keepaspectratio]{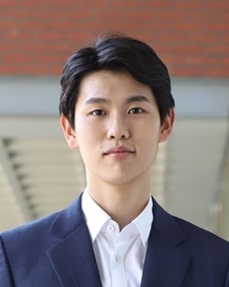}}]{Seokhyun Jeong} (Student Member, IEEE)
received the B.S. degree from the Department of Electrical and Computer Engineering, Seoul National University, Seoul, South Korea, in 2022, where he is currently pursuing the Ph.D. degree in electrical and computer engineering. His main areas of research are in artificial intelligence for wireless communications.
\end{IEEEbiography}

\begin{IEEEbiography}[{\includegraphics[width=1in,height=1.35in,clip,keepaspectratio]{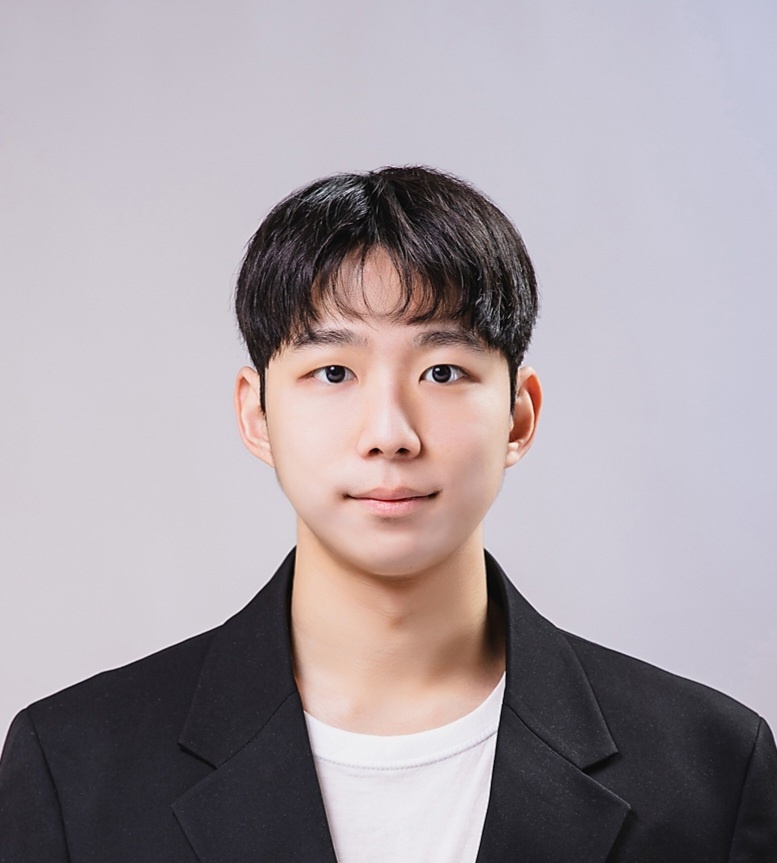}}]{Sangmok Shin} (Student Member, IEEE)
received the B.S. degree from Daegu Gyeongbuk Institute of Science and Technology (DGIST), Daegu, South Korea, in 2023.
He is currently pursuing the Ph.D. degree in Electrical and Computer Engineering at Seoul National University, Seoul, South Korea.
His research interests include AI-assisted physical-layer wireless communications, with particular emphasis on beam management, channel estimation, and channel prediction.
Mr. Shin was a recipient of the Samsung Humantech Paper Award Bronze Prize and the Graduate Student of the Year Award from the Department of Electrical and Computer Engineering in 2025.

\end{IEEEbiography}

\begin{IEEEbiography}[{\includegraphics[width=1in,height=1.35in,clip,keepaspectratio]{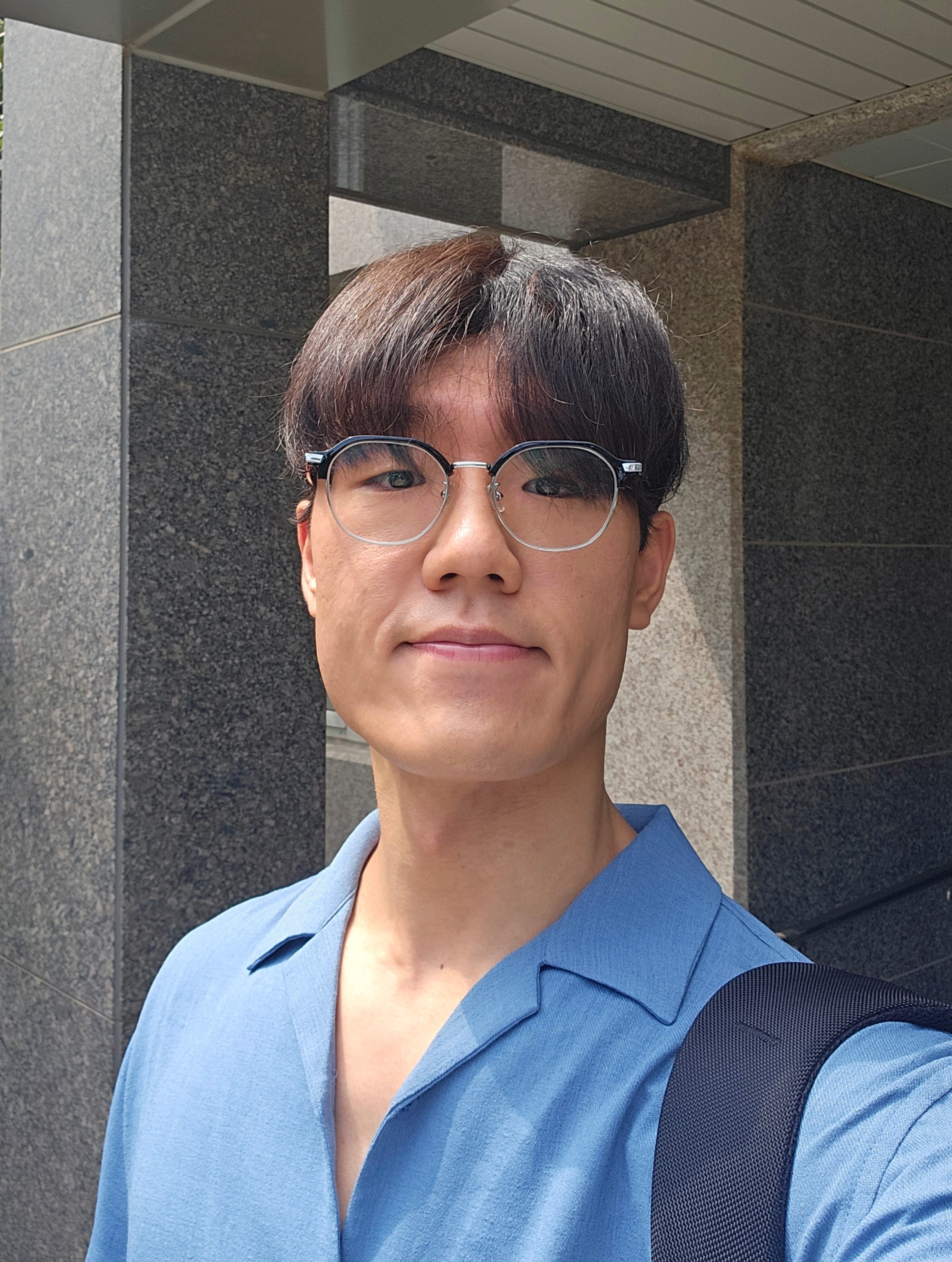}}]{Seungnyun Kim} (Member, IEEE)
received the B.S. (with honors) and Ph.D. degrees in electrical and computer engineering from Seoul National University (SNU), Seoul, South Korea, in 2016 and 2023, respectively. 

He is an Assistant Professor at the Singapore University of Technology and Design (SUTD). 
Prior to joining SUTD, he was a Postdoctoral Fellow with the Wireless Information and Network Sciences Laboratory, Massachusetts Institute of Technology, Cambridge, MA, USA. 
His research interests include information theory, optimization methods, and machine learning with applications to real-world problems, including wireless communications, network localization and navigation, and non-terrestrial networks.

Dr. Kim was a recipient of the Sejong Science Fellowship from Korean Government in 2023, the Best Ph.D. Dissertation Award from SNU in 2023, the Qualcomm Innovation Fellowship Finalist in 2021, and the Samsung Humantech Paper Award Gold Prize in 2019.
\end{IEEEbiography}

\begin{IEEEbiography}[{\includegraphics[width=1in,height=1.35in,clip,keepaspectratio]{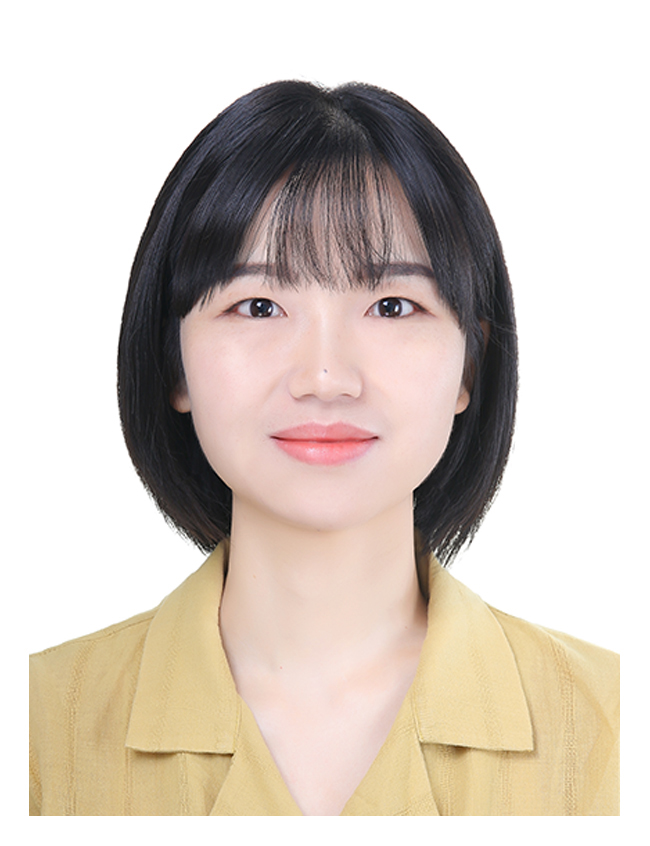}}]{Jiao Wu} (Member, IEEE)
received the B.S.\ degree in communication engineering from North China Electric Power University (NCEPU), Beijing, China, in 2015, the M.S.\ degree in electronics and communication engineering from University of Electronic Science and Technology of China (UESTC), Chengdu, China, in 2018, and the Ph.D.\ degree in electrical and computer engineering from Seoul National University (SNU), Seoul, South Korea, in 2023.

She is currently a Postdoctoral Fellow with the Computer, Electrical and Mathematical Sciences and Engineering Division (CEMSE), King Abdullah University of Science and Technology (KAUST), Thuwal, Saudi Arabia. 
Her research interests include signal processing, optimization techniques, and machine learning with the applications to reconfigurable intelligent surfaces-assisted communications, extremely large-scale antenna systems, integrated sensing and communications, and non-terrestrial networks. 
\end{IEEEbiography}

\begin{IEEEbiography}[{\includegraphics[width=1in,height=1.35in,clip,keepaspectratio]{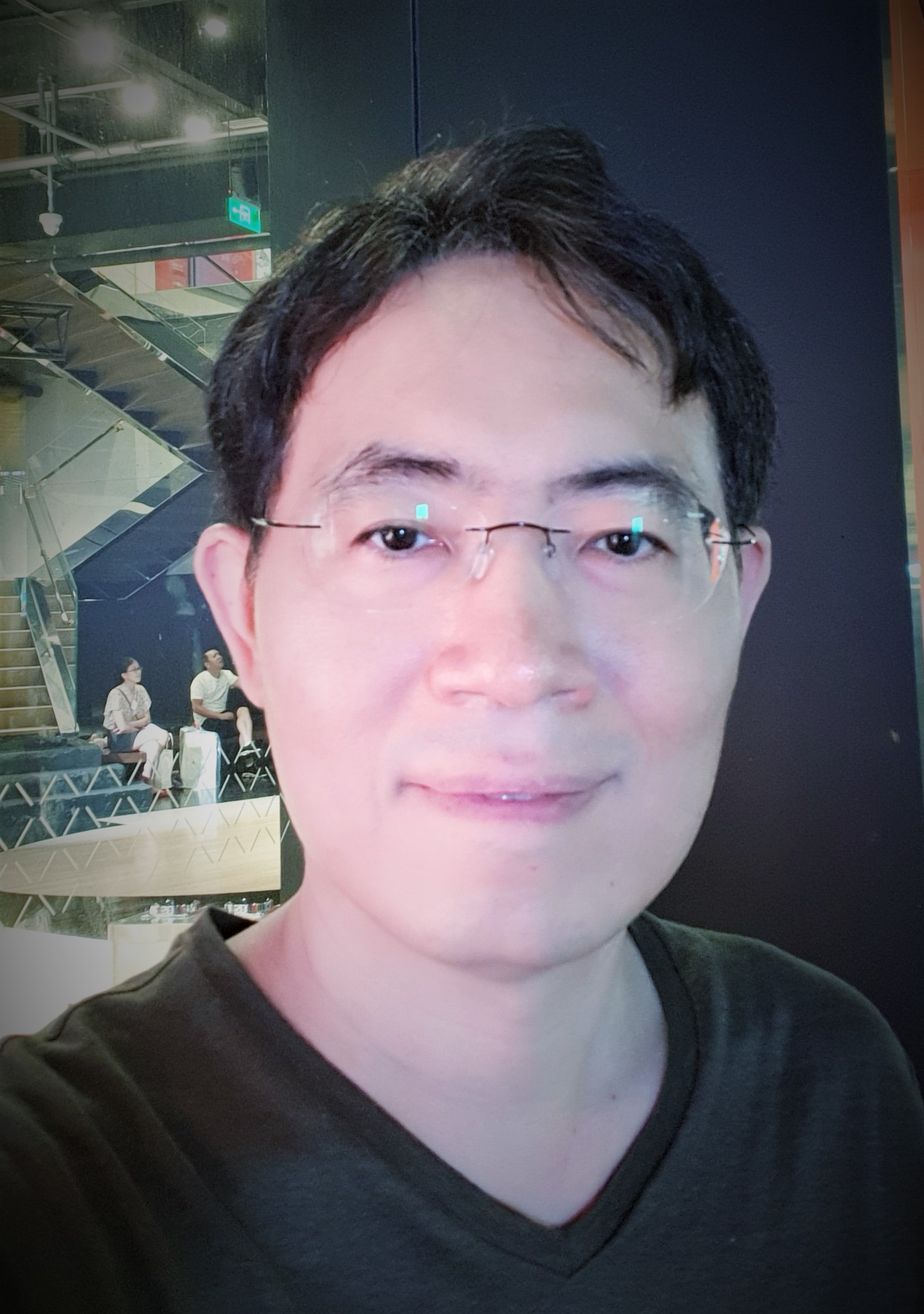}}]{Byonghyo Shim} (Fellow, IEEE) received the B.S. and M.S. degrees in Control and Instrumentation Engineering from Seoul National University, South Korea, in 1995 and 1997, respectively, and the M.S. degree in mathematics and the Ph.D. degree in Electrical and Computer Engineering from the University of Illinois at Urbana-Champaign (UIUC), Champaign, IL, USA, in 2004 and 2005, respectively. From 1997 to 2000, he was an Officer (First Lieutenant) and an Academic full-time Instructor in the Department of Electronics Engineering, Korean Air Force Academy. From 2005 to 2007, he was a Staff Engineer with Qualcomm Inc., San Diego, CA, USA. From 2007 to 2014, he was an Associate Professor with the School of Information and Communication, Korea University, Seoul. Since 2014, he has been with Seoul National University (SNU), where he is currently a Professor of the Department of Electrical and Computer Engineering and Vice Dean of Engineering College. 

His research interests include wireless communications, deep learning, and statistical signal processing. Dr. Shim was a recipient of the M. E. Van Valkenburg Research Award from the ECE Department, University of Illinois, in 2005, the Haedong Young Engineer Award from IEIE in 2010, the Irwin Jacobs Award from Qualcomm and KICS in 2016, the Shinyang Research Award from the Engineering College of SNU in 2017, the Okawa Foundation Research Award in 2020, the IEEE Comsoc AP Outstanding Paper Award in 2021, the JCN Best Paper Award in 2024, and the SNU Academic Research Award in 2025. Dr. Shim was an Elected Member of the Signal Processing for Communications and Networking (SPCOM) Technical Committee of the IEEE Signal Processing Society. He has served as an Associate Editor for IEEE Transactions on Wireless Communications (TWC), IEEE Transactions on Communications (TCOM), IEEE Transactions on Vehicular Technology (TVT), IEEE Transactions on Signal Processing (TSP), IEEE Wireless Communications Letters (WCL), and Journal of Communications and Networks (JCN) and a Guest Editor for IEEE Journal on Selected Areas in Communications (JSAC).
\end{IEEEbiography}

\end{document}